\documentclass[reprint, amsmath,amssymb, aps]{revtex4-2}

\usepackage{graphicx}
\usepackage{dcolumn}
\usepackage{bm}
\usepackage{hyperref}
\usepackage[mathlines]{lineno}
\usepackage{array}
\usepackage{longtable}
\usepackage{multirow}
\usepackage{booktabs}
\usepackage{physics}
\usepackage{float}
\usepackage{xcolor}

\begin{document}

\title{Statistical Framework for Discovery Sensitivity and Majorana Mass Estimation in \(^{136}\)Xe Neutrinoless Double Beta Decay}

\author{Pratima Singh}
\affiliation{Department of Physics, University of Lucknow, Lucknow, India}

\author{Jyotsna Singh}
\email{singh.jyotsnalu@gmail.com}
\thanks{Corresponding author}
\affiliation{Department of Physics, University of Lucknow, Lucknow, India}

\author{R.~B.~Singh}
\email{singh$_$rb@lkouniv.ac.in}
\thanks{Corresponding author}
\affiliation{Department of Physics, University of Lucknow, Lucknow, India}

\begin{abstract}
Neutrinoless double-beta decay (\(0\nu\beta\beta\)) is a sensitive probe of lepton-number violation and the Majorana nature of neutrinos. In xenon-based experiments, the expected signal rate inside the region of interest (ROI) is extremely small, requiring sensitivity estimates based on Poisson statistics and a careful treatment of detector resolution, background fluctuations, and systematic uncertainties.
\\
In this work, we develop a statistical framework relating energy resolution, ROI width, background index, isotope exposure, and discovery sensitivity for \(^{136}\)Xe-based \(0\nu\beta\beta\) experiments. The formalism combines Poisson likelihood methods with realistic background modeling and includes reconstruction-related and final-state interaction (FSI) systematic effects through an effective ROI broadening approach.
\\
Using representative detector parameters for LZ, NEXT-100, KamLAND-Zen, and nEXO, we compare expected background counts, required discovery signal statistics, and half-life sensitivities at matched exposure. The corresponding sensitivities are translated into effective Majorana mass reach within both normal- and inverted-hierarchy neutrino mass ordering. The impact of uncertainties associated with the axial-vector coupling constant \(g_A\), nuclear matrix elements, and phase-space factors is also examined.
\\
Our results show that background suppression, ROI optimization, and control of detector-related systematics are essential for extending sensitivity toward the normal-ordering regime in future \(0\nu\beta\beta\) searches.
\end{abstract}
\maketitle
\section{Introduction}
Neutrinoless double-beta decay ($0\nu\beta\beta$) is one of the most important and rare processes in the field of particle and nuclear physics. This is a hypothetical second-order weak interaction in which an even-even nucleus undergoes a transition without the emission of neutrinos \cite{Rodejohann:2011mu, Vergados:2002pv, Cremonesi:2002is}. The process can be visualised as
\begin{equation}
(A, Z) \rightarrow (A, Z+2) + 2e^-
\end{equation}
The observation of this process would establish the total lepton number violation $\Delta L=2$, B-L violation ($\Delta B=0$ and $\Delta L=2$), and at the same time it will confirm the Majorana nature of neutrinos \cite{majorana1937theory,PhysRevD.25.2951,Gomez-Cadenas:2023vca, Agostini:2022zub,Dolinski:2019nrj}. The above decay rate is commonly expressed in terms of the half-life as:
\begin{equation}
\left(T_{1/2}^{0\nu}\right)^{-1}=G^{0\nu}\,g_A^4\,|M^{0\nu}|^2\,\langle m_{\beta\beta} \rangle^2,
\label{eq:halflife}
\end{equation}
where \(G^{0\nu}\) is the phase-space factor, \(g_A\) is the axial-vector coupling constant, and \(M^{0\nu}\) is the nuclear matrix element \cite{Engel:2016xgb, Kotila:2012zza, Mirea:2015nsl}, and $\langle m_{\beta\beta} \rangle$ is the effective Majorana neutrino mass defined as
\begin{equation}
\langle m_{\beta\beta} \rangle =\left|\sum_i U_{ei}^{2} m_i\right|,
\label{eq:Majoranamass}
\end{equation}
where $U_{ei}$ are the elements of the PMNS mixing matrix and $m_i$
represent the neutrino mass eigenstates. The determination of the effective Majorana neutrino mass $\langle m_{\beta\beta} \rangle$ therefore depends strongly on neutrino oscillation parameters such as the mixing angles $\theta_{12}$, $\theta_{13}$ and the mass-squared differences $\Delta m_{21}^{2}$ and $\Delta m_{31}^{2}$. The best fit values of neutrino oscillation parameters are used to determine the effective Majorana mass in normal-hierarchy (NH) and inverted-hierarchy (IH) assumptions \cite{ParticleDataGroup:2024cfk}. But in neutrino oscillation experiments, Final State Interactions (FSI) modify the observed final-state particle distributions through scattering, absorption, and charge-exchange processes inside the nuclear medium \cite{Naaz:2018amr, Nagu:2019uco, Nagu:2019fvi, Singh:2019qac, Sharma:2023feb, Sharma:2022ype}. These effects distort the reconstructed neutrino energy and introduce systematic uncertainties in the extracted oscillation parameters. Since the effective Majorana mass is directly linked to these parameters through the PMNS matrix, FSI-induced uncertainties indirectly propagate into the estimation of
$\langle m_{\beta\beta} \rangle$ for both NH and IH neutrino mass scenarios.\\
From ~Equation\ref{eq:halflife} we observe that, measurements of the half-life directly constrains the absolute neutrino mass scale and the underlying mechanism of lepton number violation. Experimentally, the signature of $0\nu\beta\beta$ decay will be a mono-energetic peak at the transition energy $Q_{\beta\beta}$, corresponding to the summed kinetic energy of the two emitted electrons. Several isotopes are actively employed in ongoing \(0\nu\beta\beta\) decay searches, including \(^{136}\)Xe, \(^{76}\)Ge, \(^{130}\)Te, \(^{100}\)Mo, and \(^{82}\)Se. Among these, \(^{136}\)Xe has emerged as one of the most promising candidate due to its excellent scalability to large detector masses, efficient purification techniques, favorable energy resolution, and strong self-shielding capability against external radioactive backgrounds. These properties make xenon-based detectors particularly well suited for next-generation high-sensitivity \(0\nu\beta\beta\) experiments. Its decay $Q$-value which is approximately $2458~\mathrm{keV}$ lies above much of the natural radioactive background spectrum, enhancing experimental sensitivity.
The Standard Model-allowed two-neutrino double beta decay ($2\nu\beta\beta$) mode has been experimentally observed in several isotopes and it constitutes an irreducible background for $0\nu\beta\beta$ searches~\cite{Barabash}. To distinguish the neutrinoless mode, experiments rely on excellent energy resolution, ultra-low background environments, and large detector exposures.
Substantial progress in this direction has been achieved through a variety of experimental techniques employing different isotopes and detector technologies. Xenon-based experiments such as EXO-200~\cite{EXO-200:2019rkq,Yen:2016sph, Auty:2015cna}, KamLAND-Zen~\cite{KamLAND-Zen:2012mmx,Ichimura:2022kvl}, and NEXT~\cite{NEXT:2015wlq,PalmeiroPazos:2024fzd, NEXT:2023daz} have demonstrated competitive exposures and low-background performance using liquid xenon, xenon-loaded scintillator, and high-pressure gaseous xenon time projection chambers, respectively. Germanium-based experiments, including LEGEND-1000 \cite{LEGEND:2025jwu, Cattadori:2021cnk, Saleh:2026mon, Zuzel:2025lsr}, GERDA~\cite{GERDA:2020xhi,Krause:2021luk, Burlac:2021azy} and the MAJORANA Demonstrator~\cite{Majorana:2022udl}, have achieved exceptional energy resolution in searches using enriched $^{76}$Ge detectors. Similarly, experiments such as CUORE~\cite{CUORE:2021mvw} using $^{130}$Te bolometers and NEMO-3~\cite{NEMO:2008zza, NEMO-3:2016mvr} employing tracking-calorimeter techniques have provided important complementary measurements.
\\
The most stringent published half-life limit for $^{136}$Xe comes from the combined KamLAND‑Zen, which set a lower bound on the $0\nu\beta\beta$ half-life of $^{136}$Xe at $T_{1/2}^{0\nu} \sim \mathcal{O}(10^{26})~\mathrm{yr}$ corresponding to upper limits on $\langle m_{\beta\beta}\rangle$ in the range of roughly 28--122 meV \cite{KamLAND-Zen:2012mmx,KamLAND-Zen:2024eml}
\\
Current and next-generation experiments aim to probe half-lives beyond $10^{27}$--$10^{28}$ years, corresponding to effective Majorana masses in the tens of meV region. Despite substantial improvements in detector mass and background reduction, the expected number of signal events remains extremely small even for multi-year exposures in the physically relevant parameter space.
\\
In this low-count regime, statistical fluctuations in the observed event rate play a central role in data interpretation. Gaussian approximations are generally inadequate, and hence instead of this a consistent analysis which is based on Poisson statistics and likelihood-based hypothesis testing must be used. Motivated by this, we develop a unified statistical framework to accurately model signal significance, background fluctuations, and half-life sensitivity for current and next-generation $0\nu\beta\beta$ experiments.\\
In this work, the discovery sensitivity and Majorana mass estimation in $^{136}$Xe neutrinoless double beta decay are performed in three main steps. First, a statistical framework is developed to estimate the expected signal and background contributions using experimentally relevant quantities such as the isotope exposure, background index, detector efficiency, and region of interest (ROI). Second, the discovery sensitivity and half-life reach are evaluated under both background-free and background-limited conditions using Poisson-based statistical methods. Finally, the obtained half-life sensitivity
is translated into the effective Majorana neutrino mass $\langle m_{\beta\beta} \rangle$ using nuclear matrix element calculations
and neutrino oscillation parameters, while also incorporating the effect of FSI-induced uncertainties on the extraction of oscillation parameters and the corresponding propagation in normal order and inverted order mass estimates.
\section{Background Modeling of $^{136}Xe$ for $0\nu\beta\beta$}
A quantitative description of background contributions is essential for a reliable sensitivity estimate in $0\nu\beta\beta$ searches. The signal corresponds to a discrete peak at $Q_{\beta\beta}=2458~\mathrm{keV}$, whereas background processes generate continuous or line-like spectra that can populate the ROI because of finite detector resolution, partial energy deposition, or event misidentification.
The total differential background rate can be written as
\begin{equation}
B_{\mathrm{tot}}(E\mathbf{x},t)=\sum_iA_i\,S_i(E)\,P_i(\mathbf{x})\,T_i(t),
\end{equation}
where $A_i$ is the normalization of source $i$, $S_i(E)$ is its normalized energy spectrum, $P_i(\mathbf{x})$ its spatial distribution, and $T_i(t)$ is time dependence.
The expected number of background events in an ROI of width $\Delta E$ centered around $Q_{\beta\beta}$ is then
\begin{equation}
N_b=\int_{\mathrm{ROI}} dE \int_V d^3x \int dt \; B_{\mathrm{tot}}(E,\mathbf{x},t)
\simeq \mathrm{BI}\,\Delta E\,M\,t,
\label{eq:background-count}
\end{equation}
where $\mathrm{BI}$ is the background index in units of $\mathrm{counts}/(\mathrm{keV}\cdot\mathrm{kg}\cdot\mathrm{yr})$, $M$ is the active isotope mass, and $t$ is the live time.
The dominant background contributions in $^{136}$Xe experiments are as follows:
\begin{enumerate}
    \item Radioactive contamination in detector materials and nearby structures acts as an important background source. In particular, gamma rays emitted from an  isotope such as $^{214}$Bi in the $^{238}$U chain and daughters of the $^{232}$Th chain can deposit energy near $Q_{\beta\beta}$ through Compton scattering, multi-site interactions, or full absorption. These contributions are strongly detector-dependent and are often concentrated near surfaces or structural components.
    \item Cosmogenic activation can generate isotopes inside the detector medium itself. An important example for the same is $^{137}$Xe, which is produced through neutron capture by $^{136}$Xe. Since $^{137}$Xe decays by beta emission with an endpoint overlapping the signal region, it constitutes an internal and spatially broad background component \cite{Singh:2026nxh}\cite{Singh:2025gnm}.
    \item Solar neutrinos, especially originating from the $^{8}$B flux, contribute through elastic scattering processes. Although this background is irreducible, its expected contribution in the ROI is usually small compared with radiogenic and cosmogenic sources.
    \item Muon-induced processes can generate delayed backgrounds through spallation products or secondary neutrons. Even in deep underground laboratories, such effects must be included to make the background modeling more accurate.
    \item Finally,the Standard Model allowed two-neutrino double-beta decay ($2\nu\beta\beta$) of $^{136}$Xe produces a continuous spectrum extending up to the endpoint. In detectors with sufficiently good energy resolution, the leakage of this spectrum into a narrow ROI around $Q_{\beta\beta}$ is strongly suppressed and is often subdominant. Nevertheless, it remains an irreducible intrinsic background and should be included when high precision is required. Accurate background modeling is mandatory for $0\nu\beta\beta$ searches.

\end{enumerate}
\section{Poisson Statistics in the Low-Background Regime}
Searches for $0\nu\beta\beta$ decay are generally performed in an ultra-low background regime, where the expected number of background events within the ROI is of the order of unity or even smaller. In such a situation, the observed event counts are discrete and are therefore described by Poisson statistics. The Poisson probability distribution is given by
\begin{equation}
P(n\mid\mu)=\frac{\mu^n e^{-\mu}}{n!},
\end{equation}
where $n$ is the number of observed events, i.e., the actual detector counts recorded during the measurement interval within the ROI, and $\mu$ is the expected mean number of events. These observed events may originate either from the desired $0\nu\beta\beta$ signal or from background sources such as natural radioactivity, cosmic rays, or detector noise.
The observed counts fluctuate statistically about the mean value $\mu$, with standard deviation
\begin{equation}
\sigma=\sqrt{\mu}.
\end{equation}
Hence, the relative statistical fluctuation is
\begin{equation}
\frac{\sigma}{\mu}=\frac{1}{\sqrt{\mu}}.
\end{equation}
This relation clearly demonstrates that when $\mu \lesssim 1$, the fractional uncertainty becomes large, indicating that statistical fluctuations are inherently significant in the rare-event search regime.
In the large-background limit, where the expected number of background events is sufficiently high, the Poisson distribution approaches a Gaussian form. In this case, the discovery significance for a counting experiment can be approximated by
\begin{equation}
Z \approx \frac{N_s}{\sqrt{N_b}},
\label{eq:gaussian_significance}
\end{equation}
where $N_s$ is the expected number of signal events and $N_b$ is the expected number of background events. Here, $\sqrt{N_b}$ represents the standard deviation of the background counts under the background-only hypothesis. Therefore, the ratio $N_s/\sqrt{N_b}$ measures the signal excess in units of the statistical fluctuation of the background.
When the background estimate is affected by systematic uncertainties, the above expression must be modified. The expected discovery significance is then written as
\begin{equation}
Z = \frac{N_s}{\sqrt{N_b + \sigma_{N_b}^{2}}},
\end{equation}
where $\sigma_{N_b}$ denotes the standard deviation associated with the systematic uncertainty in the background prediction. The additional term accounts for the fact that the expected background is not known exactly, and both statistical and systematic uncertainties must be included to obtain a more realistic estimate of the significance.
Assuming that final-state interaction (FSI) and reconstruction effects contribute
a fractional uncertainty \(f_{\rm FSI}\) to the expected background prediction,the systematic uncertainty is written as\[\sigma_{N_b}^{\rm FSI}=f_{\rm FSI}N_b.\]
The modified discovery significance therefore becomes\[Z=\frac{N_s}{\sqrt{N_b+(f_{\rm FSI}N_b)^2}}.\]
However, this expression is only valid when $N_b \gg 1$. In the low-count regime which is relevant for $0\nu\beta\beta$ searches, it does not provide a reliable measure of significance because it ignores the discrete and asymmetric nature of Poisson fluctuations. For this reason, in the next section, the discovery significance is estimated which is based on the exact Poisson likelihood formalism.
\section{Likelihood-Based Discovery Significance}

For a given observation $n$, the Poisson likelihood under a hypothesis with mean $\mu$ is
\begin{equation}
\mathcal{L}(n\mid \mu)=\frac{\mu^n e^{-\mu}}{n!}.
\end{equation}
For background only hypothesis $\mu=N_b$ and for signal plus background hypothesis $\mu= N_s + N_b$.\\
A natural test statistic for distinguishing the background-only and signal-plus-background hypotheses is represented by the likelihood ratio of both and is represented as
\begin{equation}
\lambda = \frac{\mathcal{L}(n\mid N_b)}{\mathcal{L}(n\mid N_s+N_b)}.
\label{Poisson-eq}
\end{equation}
Substituting the Poisson form in equation~\ref{Poisson-eq}, the likelihood ratio turns to be 
\begin{equation}
\lambda = \left(\frac{N_b}{N_s+N_b}\right)^n e^{N_s},
\end{equation}
and hence
\begin{equation}
-2\ln\lambda = 2\left[n\ln\left(1+\frac{N_s}{N_b}\right)-N_s\right].
\end{equation}
For sensitivity estimates, it is customary to evaluate the median significance using the Asimov data set, in which the observed count is replaced by its expectation under the signal-plus-background hypothesis, $n=N_s+N_b$. This yields
\begin{equation}Z =\sqrt{2\left[
(N_s+N_b)\ln\left(1+\frac{N_s}{N_b}\right)-N_s\right]}
\label{eq:asimov_significance}
\end{equation}
Equation~(\ref{eq:asimov_significance}) provides the appropriate discovery significance in the rare-event regime and will be used throughout this work to estimate the signal thresholds required for evidence and discovery.
When FSI and reconstruction effects are treated as a fractional systematic
uncertainty on the expected background, we define
\[
\sigma_b = f_{\rm FSI}N_b ,
\]
where \(f_{\rm FSI}\) represents the assumed fractional uncertainty due to FSI/reconstruction effects.
The Asimov discovery significance including background uncertainty is then written as
\begin{equation}
\label{eq:asimov_FSIsignificance}
\begin{aligned}
Z_A=\Bigg[2\Bigg((N_s+N_b)\ln\!\left(\frac{(N_s+N_b)(N_b+\sigma_b^2)}{N_b^2+(N_s+N_b)\sigma_b^2}\right)\\[4pt]
-\frac{N_b^2}{\sigma_b^2}\ln\!\left(1+\frac{\sigma_b^2 N_s}
{N_b(N_b+\sigma_b^2)}\right)\Bigg)\Bigg]^{1/2}.
\end{aligned}
\end{equation}
\section{Essential Physics Ingredients for the study of $0\nu\beta\beta$ decay}
The experimental sensitivity of neutrinoless double beta decay ($0\nu\beta\beta$) searches is governed by a combination of detector energy resolution, background level, exposure, and nuclear matrix element uncertainties. For the ($0\nu\beta\beta$)decay search, the most important quantities are the \textbf{Region of Interest (ROI)}, \textbf{background scaling}, \textbf{half-life sensitivity}, and the inferred \textbf{effective Majorana mass}. These parameters collectively determine the discovery reach of a given experiment.
\subsection{Region of Interest}
The region of interest (ROI) in neutrinoless double beta decay experiments is defined as the energy interval centered around the decay Q-value \(Q_{\beta\beta}\), within which the signal is expected to appear.
Because of the finite detector energy resolution, the reconstructed signaldistribution is approximately Gaussian with standard deviation \(\sigma_E\).
A symmetric ROI is therefore commonly defined as
\begin{equation}
E \in\left[Q_{\beta\beta}-k\sigma_E,\;Q_{\beta\beta}+k\sigma_E
\right],
\end{equation}
where \(k\) determines the half-width of the analysis window.
The corresponding ROI width is
\begin{equation}
\Delta E = 2k\sigma_E
\end{equation}
When the detector resolution is quoted in terms of the full width at half maximum (FWHM), the standard deviation is related through
\begin{equation}
\sigma_E=\frac{\mathrm{FWHM}}{2.355}.
\end{equation}
Thus, the ROI may also be written as
\[\mathrm{ROI}=Q_{\beta\beta}\pm k\sigma_E.\]
The choice of ROI plays a crucial role in determining the experimental sensitivity. A narrower ROI reduces the expected background contribution, while a wider ROI increases signal acceptance at the expense of larger background fluctuations.
In the presence of FSI and reconstruction-related systematic effects, the reconstructed energy distribution may broaden beyond the ideal detector response. Such effects effectively increase the energy uncertainty and modify the ROI width. Including an additional reconstruction uncertainty term \(\sigma_{\rm FSI}\), the effective energy resolution can
be approximated as
\begin{equation}
\sigma_{\rm eff}=\sqrt{\sigma_E^2+\sigma_{\rm FSI}^2}.
\end{equation}
The corresponding effective ROI width becomes
\begin{equation}
\Delta E_{\rm eff}=2k\sigma_{\rm eff}.
\label{eq:ROIfsi}
\end{equation}
Consequently, FSI/reconstruction effects increase the effective ROI width,leading to larger background acceptance and a reduction in the achievable discovery sensitivity and half-life reach.

\subsection{Background Scaling}
Let the isotope exposure of $^{136}$Xe be defined as
\begin{equation}
\Sigma_{\mathrm{iso}} = M t,
\end{equation}
where $M$ is the mass of active $^{136}$Xe mass and $t$ is the live time. If the background spectrum is approximately flat across the ROI, then the expected number of background events scales as equation~\ref{eq:background-count}, which can be rewritten as,
\begin{equation}
N_b = \mathrm{BI}\,\Delta E\,\Sigma_{\mathrm{iso}}.
\end{equation}
where, BI is background index and $\Delta E$ is width of the ROI. from the above equation we can observe that the background count grows linearly with both exposure and ROI width. Consequently, improved energy resolution reduces background burden by reducing $\Delta E$, while material purification, shielding, event reconstruction, and topology discrimination reduces the BI.
In the presence of FSI and reconstruction-related systematic effects, the reconstructed energy distribution can broaden, effectively increasing the ROI width. Introducing an effective ROI width\(\Delta E_{\rm eff}\) expressed in Equation~\ref{eq:ROIfsi}, now the background count may be written as
\begin{equation}
N_b^{\rm eff}=\mathrm{BI}\,\Delta E_{\rm eff}\
\Sigma_{\mathrm{iso}},
\end{equation}
with
\begin{equation}
\Delta E_{\rm eff}=\Delta E(1+f_{\rm FSI}),
\end{equation}
where \(f_{\rm FSI}\) represents the fractional broadening caused by
FSI/reconstruction effects. Therefore,
\begin{equation}
N_b^{\rm eff}=N_b(1+f_{\rm FSI}),
\label{eq:backgroundfsi}
\end{equation}
demonstrating that reconstruction-related broadening increases the accepted background contribution inside the ROI and consequently degrades the overall discovery sensitivity.
\section{Half-Life Sensitivity}
\begin{figure}
    \centering
    \includegraphics[width=\linewidth]{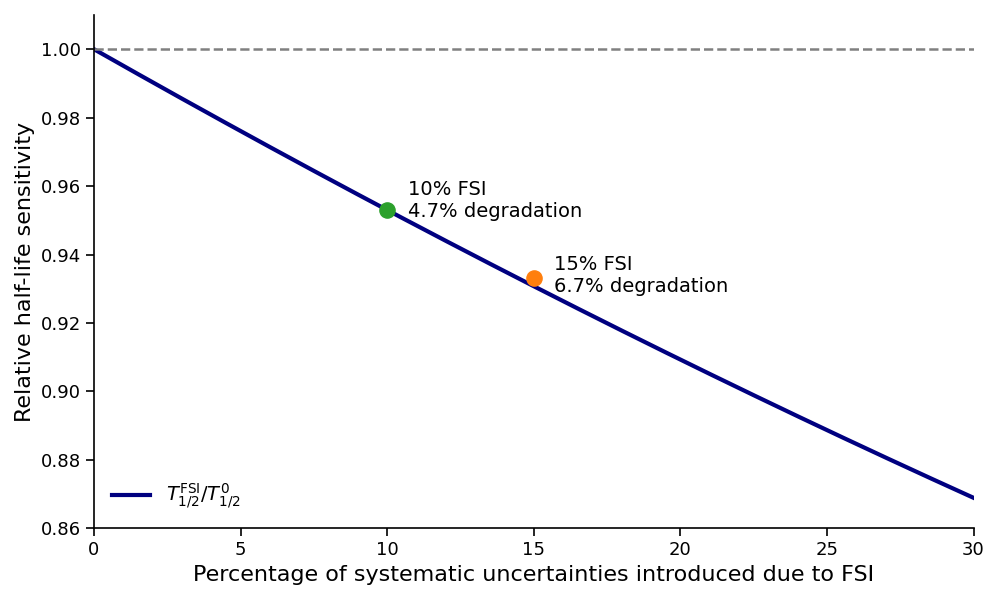}
    \caption{Relative degradation of the neutrinoless double beta decay half-life sensitivity caused by FSI.}
    \label{fig:degradation}
\end{figure}

In the background-dominated regime, the signal required for a significance threshold of $k\sigma$, scales as the square root of the statistical fluctuation of the background.
\begin{equation}
N_s^{(k\sigma)} \propto \sqrt{N_b}
\propto \sqrt{\mathrm{BI}\,\Delta E\,\Sigma_{\mathrm{iso}}}.
\end{equation}
In the presence of FSI, the reconstructed energy
distribution may broaden, leading to an effective ROI width
\(\Delta E_{\rm eff}\) expressed in Equation~\ref{eq:ROIfsi} which leads to a change in background count expressed in Equation~\ref{eq:backgroundfsi} .Therefore, in the background-dominated regime, the required signal count becomes
\begin{equation}
N_s^{(k\sigma,{\rm FSI})} \propto \sqrt{N_b^{\rm eff}}\propto
\sqrt{\mathrm{BI}\,\Delta E_{\rm eff}\,\Sigma_{\mathrm{iso}}}.
\end{equation}
and hence
\begin{equation}
N_s^{(k\sigma,{\rm FSI})}\propto
\sqrt{\mathrm{BI}\,\Delta E(1+f_{\rm FSI})\,\Sigma_{\mathrm{iso}}}.
\end{equation}
The half-life sensitivity can be derived from the expected number of
observable $0\nu\beta\beta$ signal events,
\begin{equation}
N_s =\ln 2\,\frac{N_A}{W}
\frac{a\,\varepsilon\,\Sigma_{\mathrm{iso}}}{T_{1/2}^{0\nu}},
\end{equation}
where $\Sigma_{\mathrm{iso}} = Mt$ is the isotope exposure.
Rearranging the above expression gives
\begin{equation}
T_{1/2}^{0\nu}=\ln 2\,\frac{N_A}{W}
\frac{a\,\varepsilon\,\Sigma_{\mathrm{iso }}}{N_s}.
\end{equation}
For a fixed isotope, enrichment fraction, and signal efficiency,
the quantities
\[\ln 2,\quad N_A,\quad W,\quad a,\quad \varepsilon\]
remain constant. Therefore, the half-life sensitivity scales as
\begin{equation}
T_{1/2}^{0\nu}\propto
\frac{\Sigma_{\rm iso}}{N_s}
\qquad \Longrightarrow \qquad
T_{1/2}^{0\nu,{\rm FSI}}
\propto\frac{\Sigma_{\rm iso}}{N_s^{\rm FSI}} 
\label{eq:halflife_scaling}
\end{equation}
In sensitivity calculations, the required signal count corresponding to a chosen discovery significance or confidence level is denoted by \(N_s^{\rm req}\). The presence of systematic uncertainties, particularly those arising from FSI effects, modifies the effective signal requirement and consequently alters the inferred half-life sensitivity \(T_{1/2}\). Figure~\ref{fig:degradation} illustrates the variation of the relative half-life sensitivity with increasing percentage of FSI-induced systematic uncertainty for different experimental configurations. As the FSI contribution increases, the required signal count also changes, leading to a degradation in the achievable half-life sensitivity. The modified required signal count in the presence of FSI effects is expressed as follows:
\begin{equation}
T_{1/2}^{0\nu}\propto\frac{\Sigma_{\rm iso}}{N_s^{\rm req}}
\qquad \Longrightarrow \qquad
T_{1/2}^{0\nu,{\rm FSI}}\propto\frac{\Sigma_{\rm iso}}{N_s^{\rm req,{\rm FSI}}}
\end{equation}
This relation shows that the half-life sensitivity improves with increasing isotope exposure and decreases when a larger signal threshold is required for discovery.
\subsection{Effective Majorana Mass Sensitivity}
The half-life of neutrinoless double beta decay is related to the effective Majorana neutrino mass through
\begin{equation}
\left(T_{1/2}^{0\nu}\right)^{-1}
=G^{0\nu}\,g_A^4\,|M^{0\nu}|^2\,\langle m_{\beta\beta} \rangle^2,
\end{equation}
where \(G^{0\nu}\) is the phase-space factor, \(g_A\) is the axial-vector coupling constant, and \(M^{0\nu}\) is the nuclear matrix element.
The above equation for effective Majorana mass takes the form
\begin{equation}
\langle m_{\beta\beta} \rangle =
\frac{1}{\sqrt{G^{0\nu}\,g_A^4\,|M^{0\nu}|^2\,T_{1/2}^{0\nu}}}.
\end{equation}
Thus, the experimental mass sensitivity scales as
\begin{equation}
\langle m_{\beta\beta} \rangle \propto \frac{1}{\sqrt{T_{1/2}^{0\nu}}}.
\end{equation}
Substituting the half-life sensitivity derived in the previous section, one obtains the dependence of the effective mass sensitivity on experimental parameters.\\
\\
\textbf{(i) In Background-free regime:}  
In the limit of negligible background using, \(T_{1/2}^{0\nu} \propto a\,\varepsilon\,\Sigma_{\mathrm{iso}}\), the effective Majorana neutrino mass sensitivity becomes
\begin{equation}
    \langle m_{\beta\beta} \rangle \propto
\frac{1}{\sqrt{a\,\varepsilon\, \Sigma_{iso}}}
\end{equation}
This demonstrates that the sensitivity improves with the square root of the exposure and is enhanced by higher isotopic enrichment and detection efficiency.\\
\\
\textbf{(ii) In Background-dominated regime:}  
When background contributions are non-negligible, using
\(T_{1/2}^{0\nu} \propto a\,\varepsilon
\sqrt{\Sigma_{\mathrm{iso}} / (\mathrm{BI}\,\Delta E)}\),
the effective mass sensitivity becomes
\begin{equation}
\langle m_{\beta\beta} \rangle \propto
\frac{1}{\sqrt{a\,\varepsilon}}
\left(\frac{\mathrm{BI}\,\Delta E}{\Sigma_{\mathrm{iso}}}\right)^{1/4}.
\label{eq:majorna-mass}
\end{equation}
This scaling highlights that, in the background-dominated regime, the improvement in mass sensitivity with exposure is significantly slower, following a \(\Sigma_{\mathrm{iso}}^{-1/4}\) dependence. In contrast, reductions in the background index and improvements in energy resolution play a crucial role, as the sensitivity depends on \((\mathrm{BI}\,\Delta E)^{1/4}\).
Overall, these relations emphasize that achieving competitive sensitivity to \(\langle m_{\beta\beta} \rangle\) requires not only large exposures but also stringent control of backgrounds and high energy resolution.
\section{$^{136}Xe$ Experimental Inputs In $0\nu\beta\beta$ decay physics}
Experiments based on the $^{136}Xe$ isotope plays a crucial role in the search of $0\nu\beta\beta$ decay. The attractiveness of $^{136}Xe$ lies in high Q-value $(\sim 2458 keV)$, which places the signal above most natural radioactive backgrounds. Current and next generation experiments exploit different detection techniques. KamLand-Zen uses xenon loaded liquid scintillator to achieve large source mass, NEXT employs high pressure gaseous xenon time projection chamber to obtain better energy resolution and topological event discrimination, and nEXO aims to combine large mass with improved background control in Liquid Xenon TPC. Although, LUX-ZEPLIN is not optimized for $0\nu\beta\beta$ experiment, but this experiment provides valuable constraints xenon experiment. Together these experimental results will help probe the better half life of $0\nu\beta\beta$ decay and effective Majorana masses $<m_{\beta\beta}>$.
\\
The experiments considered in this work are discussed below.
\subsection{LUX-ZEPLIN (LZ) Experiment}
The LZ experiment is a two-phase xenon time projection chamber designed for search of weakly interacting massive particles (WIMPs), but it is also sensitive for $0\nu\beta\beta$ decay search. It utilizes 7 tonnes of liquid xenon (LXe). The natural abundance of $\rm ^{136}Xe$ is 8.89\%, yielding 623 kg of $\rm ^{136}Xe$. 
\\
For the LZ-based interpretation considered here, the estimate background for the experiment is taken from the inner fiducial region used in the published study of $0\nu\beta\beta$ sensitivity \cite{LZ:2019qdm}. The fiducial selection
\[26 < z < 96~\mathrm{cm}, \qquad r < 39~\mathrm{cm},\]
corresponds to an effective liquid xenon mass of approximately $967~\mathrm{kg}$.
The analysis assumes an energy resolution of $1.0\%$ at $Q_{\beta\beta}$ and uses the ROI
\[2433.3~\mathrm{keV} < E_{\mathrm{dep}} < 2482.4~\mathrm{keV}.\]
Taking $Q_{\beta\beta}=2457.83~\mathrm{keV}$, we estimate
\begin{equation}
\sigma_E^{\mathrm{LZ}} = 0.01 \times 2457.83 \approx
24.58~\mathrm{keV},
\end{equation}
so that the full $\pm1\sigma$ ROI width is
\begin{equation}
\Delta E_{\mathrm{LZ}} \approx 2\sigma_E \approx 49.16~\mathrm{keV}.
\end{equation}
Assuming natural xenon with $^{136}$Xe abundance $(a=0.089)$, the corresponding active isotope mass in this region will be
\begin{equation}
M_{136}^{\mathrm{LZ}} = 0.089 \times 967 \approx 86.1~\mathrm{kg}.
\end{equation}
Using the published total background count $(N_b)$ of $35.6$ events for $1000$ days \cite{LZ:2019qdm}, the isotope-normalized background index is estimated using Equation~\ref{eq:background-count}
\begin{equation}
\begin{aligned}
\mathrm{BI}_{\mathrm{LZ}}=
\frac{35.6}{49.16 \times 86.1 \times 2.738}
\approx \\
3.07\times 10^{-3}~\mathrm{counts}/(\mathrm{keV}\cdot\mathrm{kg}\cdot\mathrm{yr}).
\end{aligned}
\end{equation}
This $\mathrm {BI_{LZ}}$ will be used to estimate the sensitivity of the experiment in $1\sigma$ range
\subsection{NEXT-100 Experiment}
The NEXT experiment is dedicated to search for $0\nu\beta\beta$ in $^{136} \mathrm Xe$ using an electroluminescent time projection chamber. It uses 100 kg of xenon gas enriched to 91\% in $^{136} \mathrm Xe$. The experiment is designed to achieve excellent calorimetric resolution together with topological background discrimination. For the NEXT-100 experiment, detector performance parameters and background assumptions used in this work are adopted from the sensitivity analysis reported in \cite{NEXT:2015wlq}. 
The quoted background index in the paper \cite{NEXT:2015wlq} is
\begin{equation}
\mathrm{BI}_{\mathrm{NEXT}}=4\times10^{-4}
~\mathrm{counts}/(\mathrm{keV}\cdot\mathrm{kg}\cdot\mathrm{yr}).
\end{equation}
For the present analysis, an effective energy window of
\[\Delta E_{\mathrm{NEXT}} = 17~\mathrm{keV}\] corresponding to the \(\pm1\sigma\) region around \(Q_{\beta\beta}\) is considered.

In this paper the expected energy resolution at the double-beta decay endpoint is reported as \(0.5\%\) FWHM at $Q_{\beta\beta}$
Assuming a Gaussian response, the relation between FWHM and standard deviation is
\begin{equation}
\mathrm{FWHM}=2.355\,\sigma_E.
\end{equation}
Therefore, the fractional Gaussian resolution becomes
\begin{equation}
\frac{\sigma_E}{E}
=\frac{\frac{FWHM}{E}}{2.355}\approx2.12\times10^{-3}=0.212\%.
\end{equation}
Thus, the absolute energy spread at \(Q_{\beta\beta}\) is
\begin{equation}
\sigma_E^{\mathrm{NEXT}}=0.00212 \times 2457.83\approx 5.21~\mathrm{keV}.
\end{equation}
so that the full $\pm1\sigma$ ROI width is
\begin{equation}
\Delta E \approx 2\sigma_E \approx 10.42~\mathrm{keV}.
\end{equation}
This narrow ROI is one of the main advantages of gaseous xenon TPC technology, as it strongly suppresses background leakage near \(Q_{\beta\beta}\). These experimental inputs will be used in this work.
\subsection{KamLAND-Zen Experiment}
The KamLAND-Zen experiment searches for neutrinoless double-beta decay ($0\nu\beta\beta$) in $^{136}\mathrm{Xe}$ using xenon-loaded liquid scintillator deployed in the KamLAND detector. The upgraded KamLAND-Zen 800 employs approximately $745~\mathrm{kg}$ of xenon enriched in $^{136}\mathrm{Xe}$ and this improved purification helps to suppress radioactive backgrounds near the region of interest. Using KamLAND-Zen 800 data set, which accumulated a total exposure of $2.097~\mathrm{ton\cdot yr}$ of $^{136}$Xe. Combining this with earlier data, a lower limit on the half-life was reported as $T_{1/2}^{0\nu} > 3.8 \times 10^{26}~\mathrm{yr}$ at $90\%$ confidence level \cite{KamLAND-Zen:2024eml}. For the KamLAND-Zen experiment, the detector performance and background estimates are taken from the latest published results \cite{KamLAND-Zen:2024eml}. 
For the present comparison, we consider a reduced fiducial exposure of
\begin{equation}
\Sigma_{\mathrm{KLZ}} = 1.13~\mathrm{ton\cdot yr} = 1130~\mathrm{kg\cdot yr},
\end{equation}
together with the published analysis energy window
\begin{equation}
2.35 < E < 2.70~\mathrm{MeV},
\end{equation}
which corresponds to an effective ROI width of
\begin{equation}
\Delta E_{\mathrm{KLZ}} = 350~\mathrm{keV}.
\end{equation}
Using a representative total background count of
\begin{equation}
N_b^{\mathrm{KLZ}} = 60.88,
\end{equation}
the corresponding background index is estimated as
\begin{equation}
\mathrm{BI}_{\mathrm{KLZ}}=
\frac{60.88}{350 \times 1130}
\approx 1.54\times 10^{-4}~\mathrm{counts}/(\mathrm{keV}\cdot\mathrm{kg}\cdot\mathrm{yr}).
\end{equation}
\subsection{nEXO  Experiment}
The nEXO experiment is a tonne-scale search for \(0\nu\beta\beta\) using liquid xenon (LXe) enriched in \(^{136}\mathrm{Xe}\). The detector consists of a large liquid xenon time projection chamber (TPC) containing approximately \(5000~\mathrm{kg}\) of xenon enriched to \(90\%\) in \(^{136}\mathrm{Xe}\). The detector design is optimized to achieve ultra-low radioactive background levels together with excellent energy resolution \cite{nEXO:2021ujk}.
\\
For the present study, the detector response and background assumptions are adopted from the updated nEXO sensitivity analysis reported ~\cite{nEXO:2021ujk}. The expected background index is quoted as
\begin{equation}
\mathrm{BI}=7\times10^{-5}
~\mathrm{counts}/(\mathrm{FWHM}\cdot\mathrm{kg}\cdot\mathrm{yr}),
\end{equation}
where FWHM denotes the full width at half maximum of the detector energy response at the \(0\nu\beta\beta\) \(Q\)-value. The projected fractional Gaussian energy resolution at the double-beta decay endpoint is stated as \cite{nEXO:2021ujk}
\begin{equation}
\frac{\sigma_E}{E}\simeq 0.8\%\qquad\text{at}\qquad
Q_{\beta\beta}=2457.83~\mathrm{keV}.
\end{equation}
Thus, the absolute energy spread at the endpoint becomes
\begin{equation}
\sigma_E^{\mathrm{nEXO}}=0.008 \times 2457.83\approx19.66~\mathrm{keV}.
\end{equation}
so that the full $\pm1\sigma$ ROI width is
\begin{equation}
\Delta E_{\mathrm{nEXO}}\approx2\sigma_E\approx39.3~\mathrm{keV}.
\end{equation}
and the corresponding full width at half maximum is
\begin{equation}
\mathrm{FWHM}=2.355\times19.66\approx46.3~\mathrm{keV}.
\end{equation}
Converting the quoted background index in equation (48) into the conventional units of counts per keV per kg per year gives
\begin{equation}
\mathrm{BI}_{\mathrm{nEXO}}=\frac{7\times10^{-5}}{46.3}
\approx 1.51\times10^{-6}
~\mathrm{counts}/(\mathrm{keV}\cdot\mathrm{kg}\cdot\mathrm{yr}).
\end{equation}
This exceptionally low background level, combined with ton-scale isotope mass, gives nEXO one of the strongest projected sensitivities to neutrinoless double beta decay among next-generation experiments.
\section{Background Comparison of different experiments at Fixed \(^{136}\)Xe Isotope Exposure}
To compare the experiments on equal footing, we choose a common isotope exposure
\begin{equation}
\Sigma_{\mathrm{iso}} = 100~\mathrm{kg\cdot yr}
\end{equation}
of $^{136}$Xe for all the experiments. The expected background count for different experiments with and without FSI is calculated at fixed exposure and is shown in Table ~\ref{tab:NbROI}\\

\begin{table}[htbp]

\centering
\caption{Effective background counts in ROI region without FSI systematic and with FSI systematic that induce ROI broadening for different \(^{136}\)Xe-based experiments.}
\label{tab:NbROI}

\begin{tabular}{lccc}
\hline
\hline

Experiment &
\(N_b\) &
\(N_b\) (\(10\%\) FSI) &
\(N_b\) (\(15\%\) FSI)
\\
\hline
LZ
& 15.10
& 16.61
& 17.37
\\
NEXT-100
& 0.68
& 0.748
& 0.782
\\
KamLAND-Zen
& 5.39
& 5.93
& 6.20
\\
nEXO
& 0.059
& 0.065
& 0.068
\\
\hline
\hline
\end{tabular}
\end{table}
The comparison highlights the critical role of background reduction in future \(0\nu\beta\beta\) searches. Since the discovery sensitivity scales approximately as \(N_s \propto \sqrt{N_b}\) in the background-dominated regime, even moderate reductions in background can lead to substantial improvements in half-life sensitivity and effective Majorana mass reach.This becomes especially important for probing the normal-hierarchy region, where the expected signal rate is extremely small.\\
FSI related systematic effects can further modify the effective background contribution by broadening the reconstructed energy distribution and increasing event misidentification within the ROI. Consequently, experiments with superior reconstruction capabilities and lower systematic uncertainties are expected to provide stronger sensitivity to rare \(0\nu\beta\beta\) signals.

\section{Approximate Discovery Thresholds of different \(^{136}\)Xe based experiments}
For rough scaling estimates, one may employ the Gaussian approximation already introduced in Equation~(\ref{eq:gaussian_significance})
\begin{equation}
Z \approx \frac{N_s}{\sqrt{N_b}}.
\end{equation}
This relation provides a simple scaling between the required signal yield and the expected background, leading to
\begin{equation}
N_s^{(3\sigma)} \approx 3\sqrt{N_b},
\qquad
N_s^{(5\sigma)} \approx 5\sqrt{N_b}.
\end{equation}
These values should only be interpreted as approximate scaling benchmarks. Rigorous evidence and discovery thresholds in the rare-event regime must be obtained from the exact Poisson-based significance formula expressed in Equation~(\ref{eq:asimov_significance}) and Equation~(\ref{eq:asimov_FSIsignificance}).
\begin{table}[htbp]
\scriptsize
\setlength{\tabcolsep}{3pt}
\caption{Required signal counts \(N_s^{\rm req}\) for different discovery significances. FSI columns show required signal counts at \(10\%\) and \(15\%\) systematic uncertainties respectively.}
\label{tab:compact_significance}
\begin{ruledtabular}
\begin{tabular}{ccccc}
Exp. &
Sig. &
Gaussian &
Poisson &
FSI (\(10\%\) / \(15\%\))
\\
\hline
LZ
& \(3\sigma\)
& 11.66
& 13.07
& 13.59 / 14.42
\\
& \(5\sigma\)
& 19.43
& 23.25
& 24.16 / 25.63
\\
NEXT-100
& \(3\sigma\)
& 2.47
& 3.72
& 3.85 / 3.91
\\
& \(5\sigma\)
& 4.12
& 7.36
& 7.58 / 7.69
\\
KamLAND-Zen
& \(3\sigma\)
& 6.97
& 8.34
& 8.66 / 9.16
\\
& \(5\sigma\)
& 11.61
& 15.27
& 16.10 / 17.34
\\
nEXO
& \(3\sigma\)
& 0.73
& 1.77
& 1.78 / 1.80
\\
& \(5\sigma\)
& 1.21
& 3.84
& 3.89 / 3.97
\\
\end{tabular}
\end{ruledtabular}
\end{table}
Table~\ref{tab:compact_significance} compares the required signal counts for different discovery significances using Gaussian, Poisson, and FSI-modified statistical treatments. 
The inclusion of \(10\%\) and \(15\%\) FSI systematic uncertainties in Poissons Discovery significance increases the required signal counts, particularly for experiments with larger expected background levels such as LZ and KamLAND-Zen. 
\begin{figure}[htbp]
\centering
\includegraphics[width=\linewidth]{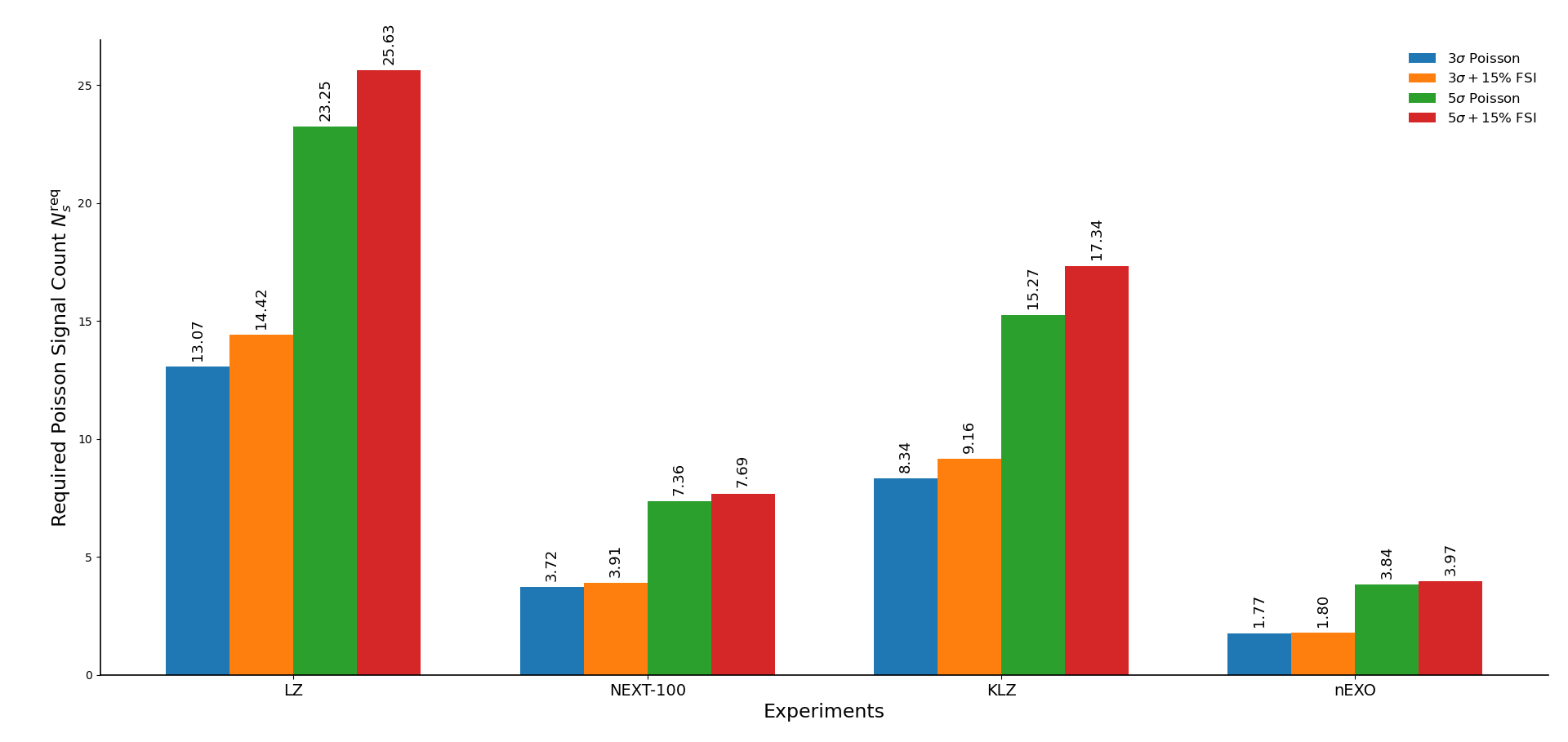}
\caption{Required signal counts \(N_s^{\rm req}\) for \(3\sigma\) and \(5\sigma\) discovery sensitivities with and without \(15\%\) FSI systematic uncertainty. The inclusion of FSI effects increases the signal requirement, particularly for higher-background experiments.}
\label{fig:PoissonNrequiredandfsi}
\end{figure}
In contrast,  as shown in Figure~\ref{fig:PoissonNrequiredandfsi} low background experiments i.e. nEXO exhibit comparatively small sensitivity degradation due to inclusion of FSI, demonstrating the importance of background suppression for future \(0\nu\beta\beta\) searches.\\
Since the half-life sensitivity scales as $T_{1/2}^{0\nu}\propto 1/N_s^{\rm req}$,hence after the inclusion of FSI systematic uncertainty there will be fractional sensitivity loss due to this systematics and that can be calculated using $\left(1-N_s^{0}/N_s^{\rm FSI}\right)\times100$, where $N_s^{0}$ corresponds to the nominal case without FSI effects and $N_s^{\rm FSI}$ includes the effective 10\% and 15\% reconstruction broadening.
\begin{figure}[htbp]
    \centering
    \includegraphics[width=\linewidth]{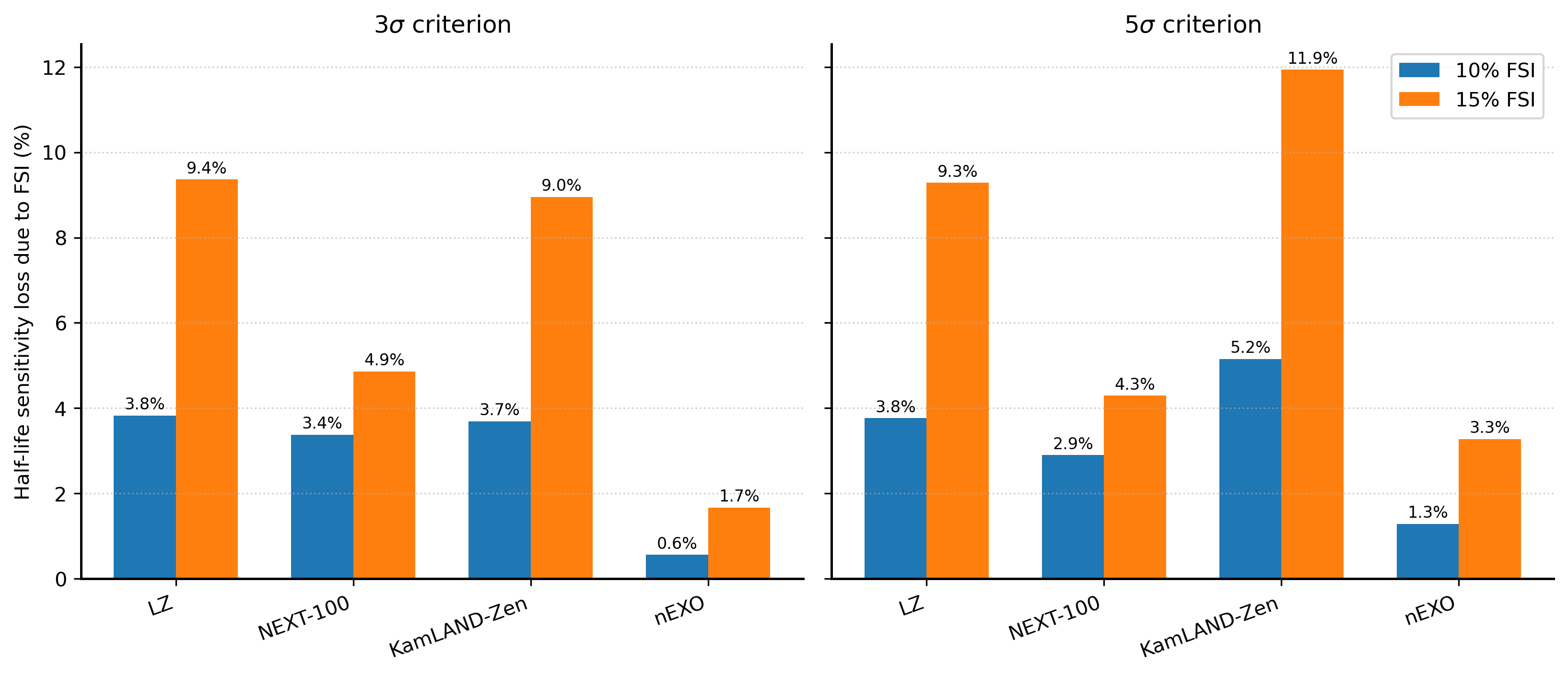}
    \caption{Percentage degradation in the neutrinoless double beta decay half-life sensitivity caused by FSI induced systematic error for different $^{136}$Xe-based experiments.}
    \label{visible_fsi_degradation}
\end{figure}
The Figure~\ref{visible_fsi_degradation} illustrates that the impact of FSI effects varies across experiments depending on their statistical and background conditions.

\section{Experimental Sensitivity to the Effective Majorana Mass}
\label{sec:Experiment}
In order to compare the oscillation-derived Majorana mass regions with the reach of present and future $^{136}$Xe-based experiments, we estimate the effective Majorana neutrino mass sensitivity for LZ, NEXT-100, KamLAND-Zen, and nEXO using the exposure--background framework developed in this work. The purpose of this analysis is to determine how close the current experimental sensitivity lies to the IH and NH regions predicted by neutrino oscillation data.
The decay rate for neutrinoless double beta decay is given in Equation~\eqref{eq:halflife}. and the effective Majorana neutrino mass defined in Equation~\eqref{eq:Majoranamass}.
These relations implies that the effective Majorana mass sensitivity scales inversely with the square root of the half-life sensitivity,
\begin{equation}
m_{\beta\beta}
\propto
\frac{1}{\sqrt{T_{1/2}^{0\nu}}}.
\end{equation}
Experimentally, the half-life sensitivity depends on the isotope exposure and the required signal threshold. For a fixed confidence level or discovery criterion, the half-life sensitivity approximately scales as expressed in Equation~\eqref{eq:halflife_scaling}.
Combining these relations gives the scaling behavior of the effective Majorana mass sensitivity;
\begin{equation}
m_{\beta\beta}^{\rm sens}=m_{\beta\beta}^{\rm ref}
\sqrt{\frac{N_s^{\rm req}}{N_{\rm ref}}
\frac{\Sigma_{\rm ref}}{\Sigma_{\rm exp}}},
\end{equation}
where $(m_{\beta\beta}^{\rm ref},\Sigma_{\rm ref})$ represent the reference Majorana mass sensitivity and reference isotope exposure.
For normalization, we adopt the benchmark sensitivity
\[m_{\beta\beta}^{\rm ref}=20~\mathrm{meV},\]
which approximately corresponds to the lower part of the inverted-hierarchy region commonly used in $^{136}$Xe sensitivity studies. The associated reference exposure is taken as
\[\Sigma_{\rm ref}=0.83~\mathrm{ton\cdot yr},\]
representing a typical exposure scale for next-generation xenon-based experiments. The reference signal threshold
\[
N_{\rm ref}=0.69
\]
is adopted from the background-free Poisson discovery criterion discussed in Ref.~\cite{Singh:2019vyc}. In the zero-background limit, the probability of observing at least one signal event is given by
\begin{equation}
P(n\ge1)=1-e^{-N_s}.
\end{equation}
Requiring a median discovery probability of \(50\%\),
\begin{equation}
P(n\ge1)=0.5,
\end{equation}
gives
\begin{equation}
N_{\rm ref}=\ln2\approx0.69.
\end{equation}
This value is therefore used as a conventional normalization point corresponding to the median discovery threshold in the background-free regime. In the present work, realistic background-dependent discovery sensitivities are obtained through the Poisson significance calculation, while \(N_{\rm ref}=0.69\) is retained only as the reference normalization constant in the scaling relation. The effective Majorana neutrino mass at $3\sigma$ and $5\sigma$ confidence levels for various experiments are summarized in Table~\ref{tab:maj_mass}.
\begin{table}[htbp]
\caption{
Effective Majorana mass sensitivities for different \(^{136}\)Xe-based experiments
at \(3\sigma\) and \(5\sigma\) confidence levels, with and without
\(15\%\) FSI effects.
}
\label{tab:maj_mass}
\centering

\begin{ruledtabular}
\begin{tabular}{lcccc}

Exp.
& \(3\sigma\)
& \(3\sigma\,+\)FSI
& \(5\sigma\)
& \(5\sigma\,+\)FSI
\\

\hline

LZ
& 163.2
& 171.5
& 217.7
& 228.6
\\

NEXT
& 80.68
& 83.47
& 113.48
& 115.99
\\

KLZ
& 59.6
& 62.45
& 80.6
& 85.9
\\

nEXO
& 92.3
& 93.1
& 136.1
& 138.2
\\

\end{tabular}
\end{ruledtabular}

\end{table}

The obtained sensitivities can now be compared with the oscillation-derived Majorana mass regions. Current global oscillation fits indicate that the inverted-hierarchy region corresponds approximately to
\[\langle m_{\beta\beta}\rangle_{\rm IH}\sim15\text{--}50~\mathrm{meV},\]
while the normal-hierarchy region lies near
\[\langle m_{\beta\beta}\rangle_{\rm NH}\sim1\text{--}10~\mathrm{meV}.\]
The determination of the effective Majorana neutrino mass
$\langle m_{\beta\beta} \rangle$ in IH and NH depends on neutrino oscillation parameters. In neutrino oscillation experiments, FSI modify the observed final-state particle distributions through scattering, absorption, and charge-exchange processes inside the nuclear medium. These effects distort the reconstructed neutrino energy and introduce systematic uncertainties in the extracted oscillation parameters. Since the effective Majorana mass is directly related to these parameters through the PMNS mixing matrix, FSI-induced uncertainties indirectly propagate into the estimation of $\langle m_{\beta\beta} \rangle$ for both normal-hierarchy and inverted-hierarchy neutrino mass scenarios.
The present estimates indicate that none of the considered experimental configurations has yet reached the generic normal-hierarchy parameter space. Even the projected nEXO sensitivity remains above the lower IH boundary,although it approaches the upper part of the IH region more closely than current experiments.
Among the experiments considered here, NEXT-100 and nEXO benefit from improved background rejection and superior energy resolution, which reduce the required signal threshold and improve the corresponding Majorana mass reach. In contrast, LZ is primarily optimized for dark matter searches rather than neutrinoless double beta decay, leading to larger effective background contributions and weaker $0\nu\beta\beta$ sensitivity despite its large xenon
mass.
These results demonstrate that future progress toward probing the
normal-ordering region will require simultaneous improvement in isotope exposure, background suppression, detector energy resolution, reconstruction algorithms, and nuclear matrix element calculations.
\section{$^{136}$Xe Neutrinoless Double Beta Decay and Majorana Mass Sensitivity}
\subsection{Majorana Mass Extraction and Theoretical Uncertainties}
The interpretation of neutrinoless double beta decay ($0\nu\beta\beta$) experiments in terms of the effective Majorana neutrino mass is based on the standard light-neutrino exchange mechanism. The relation between the half-life sensitivity and the effective Majorana mass is given by
\begin{equation}
\langle m_{\beta\beta}\rangle=\frac{m_e}
{g_A^2 |M^{0\nu}| \sqrt{G^{0\nu} T_{1/2}^{0\nu}}},
\label{eq:maj_mass_xe}
\end{equation}
where $m_e$ is the electron mass,$G^{0\nu}$ is the phase-space factor,$M^{0\nu}$ is the nuclear matrix element (NME),
and $g_A$ is the axial-vector coupling constant.
Equation~\eqref{eq:maj_mass_xe} shows that the extracted Majorana mass depends not only on the experimental half-life sensitivity but also on theoretical quantities associated with nuclear structure and weak interaction effects. Consequently, even when the experimental half-life limit is precisely known, the inferred neutrino mass can still exhibit substantial theoretical uncertainty.
For $^{136}$Xe, the phase-space factor is commonly taken as
\begin{equation}
G^{0\nu}=3.56\times10^{-14}\ {\rm yr}^{-1}.
\end{equation}
The dominant theoretical uncertainty originates from the nuclear matrix element calculations. Different many-body nuclear approaches such as QRPA, IBM-2, interacting shell model (ISM), energy density functional (EDF), generator coordinate method (GCM), and covariant density functionaltheory (CDFT) predict substantially different values of $M^{0\nu}$.
The presently available calculations span approximately
\begin{equation}
1.1 \le |M^{0\nu}| \le 4.77.
\end{equation}
Table~\ref{tab:nme_methods} summarizes representative NME calculations for $^{136}$Xe obtained using different theoretical approaches.
\begin{table}[h!]
\centering
\caption{Representative nuclear matrix element calculations for
$^{136}$Xe neutrinoless double beta decay
\cite{REBEIRO2020135702}.
}
\label{tab:nme_methods}
\begin{tabular}{lc}
\hline
Method & $M^{0\nu}$ \\
\hline
Deformed WS-QRPA & 1.11 \\
pnQRPA & 2.91 \\
Skyrme-HFB-QRPA & 1.55 \\
Spherical QRPA & 2.46 \\
ISM (Strasbourg--Madrid) & 2.19 \\
ISM (Michigan) & 1.46 \\
CDFT & 4.24 \\
NREDF & 4.77 \\
IBM-2 & 3.05 \\
GCM & 2.35 \\
\hline
\end{tabular}
\end{table}
A second major source of uncertainty arises from the effective axial-vector coupling constant $g_A$. While the free nucleon value is
\[g_A = 1.27,\]nuclear medium effects, many-body correlations, and limited model spaces can lead to quenching of the effective coupling inside nuclei. Present estimates typically span the interval
\begin{equation}
0.6 \le g_A \le 1.27.
\end{equation}
Since the Majorana mass scales as
\begin{equation}
\langle m_{\beta\beta}\rangle\propto g_A^{-2},
\end{equation}
even moderate quenching substantially weakens the inferred neutrino mass sensitivity.
The interpretation of $0\nu\beta\beta$ results also depends on the
underlying decay mechanism. Throughout this work, the standard light
Majorana neutrino exchange scenario is assumed. However, alternative
mechanisms involving heavy neutrino exchange, right-handed weak currents, supersymmetric interactions, or other lepton-number-violating operators can modify both the half-life expression and the extracted particle
physics parameters.
\subsection{Majorana Mass Range from Nuclear and Axial-Coupling Uncertainties}
For a fixed half-life sensitivity, the minimum inferred Majorana mass is obtained using the largest NME and largest axial coupling, whereas the maximum mass corresponds to the smallest NME and strongest quenching:
\begin{equation}
\langle m_{\beta\beta}\rangle_{\min}=
\frac{m_e}{g_{A,\max}^2|M^{0\nu}|_{\max}\sqrt{G^{0\nu}T_{1/2}^{0\nu}
}},
\end{equation}
and
\begin{equation}
\langle m_{\beta\beta}\rangle_{\max}
=\frac{m_e}{g_{A,\min}^2|M^{0\nu}|_{\min}
\sqrt{G^{0\nu}T_{1/2}^{0\nu}}}.
\end{equation}
Therefore, the effective Majorana mass lies within the interval
\begin{equation}
\begin{aligned}
\frac{m_e}{g_{A,\max}^2|M^{0\nu}|_{\max}\sqrt{
G^{0\nu}T_{1/2}^{0\nu}}}
\le\langle m_{\beta\beta}\rangle \\
\le\frac{m_e}{g_{A,\min}^2|M^{0\nu}|_{\min}
\sqrt{G^{0\nu}T_{1/2}^{0\nu}}}.
\end{aligned}
\end{equation}
This equation can also be written as
\begin{equation}
\frac{C}
{g_{A,\max}^2|M^{0\nu}|_{\max}}
\le\langle m_{\beta\beta}\rangle\le
\frac{C}
{g_{A,\min}^2|M^{0\nu}|_{\min}}.
\label{eq:nmega}
\end{equation}
Where the constant C depends on experimental half-life senstivity.
From the above equation we can observe that uncertainties in nuclear structure and axial-current renormalization propagate directly into the inferred neutrino Majorana mass limits.
\subsection{Experimental Majorana Mass Sensitivity}
\begin{figure}[H]
    \centering
 \includegraphics[width=1\linewidth]{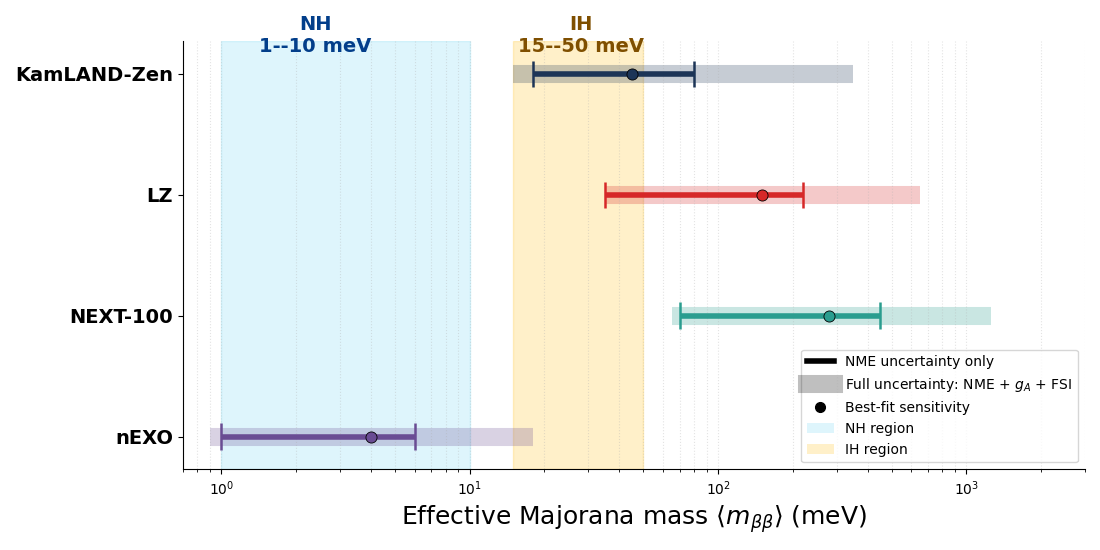}
    \caption{Dependence of the effective Majorana neutrino mass sensitivity on the nuclear matrix element $|M^{0\nu}|$ and axial coupling constant $g_A$ for different $^{136}$Xe-based experiments. The shaded regions indicate the approximate normal-hierarchy (NH) and inverted-hierarchy (IH) neutrino mass bands inferred from global oscillation data.}
    \label{fig:nmega}
\end{figure}

Using the reported and projected half-life sensitivities for different $^{136}$Xe-based experiments, the corresponding Majorana mass ranges are estimated by varying both the NME and axial coupling within the intervals discussed above.
For convenience, the extracted Majorana mass can be expressed in the form Eq.~\eqref{eq:nmega}
Using the available experimental and projected half-life limits, the
value of corresponding coefficients C is estimated as
\begin{align}
{\rm EXO\mbox{-}200}: \qquad & C = 677~{\rm meV},
\\
{\rm NEXT\mbox{-}100}: \qquad & C = 507~{\rm meV},
\\
{\rm LZ}: \qquad & C = 263~{\rm meV},
\\
{\rm KamLAND\mbox{-}Zen}: \qquad & C = 139~{\rm meV},
\\
{\rm nEXO}: \qquad & C = 7.37~{\rm meV}.
\end{align}
The resulting Majorana mass ranges for different experiments and different values of $g_A$ are summarized in Table~\ref{tab:majorana_ga_C} and expressed in Figure~\ref{fig:nmega}.
\begin{table*}[t]
\caption{Effective Majorana mass sensitivity for different $^{136}$Xe experiments as a function of the axial coupling $g_A$. The mass intervals are obtained by varying the nuclear matrix element over $1.1 \le |M^{0\nu}| \le 4.77$.}
\label{tab:majorana_ga_C}
\begin{ruledtabular}
\begin{tabular}{ccccc}
Experiment & $T_{1/2}^{0\nu}$ (yr) & $C$ (meV) & $g_A$ &
$\langle m_{\beta\beta}\rangle$ (meV) \\
\hline
KamLAND-Zen
& $>3.8\times10^{26}$
& 139
& 1.27
& 18--79 \\

&
&
& 1.00
& 29--126 \\

&
&
& 0.80
& 46--198 \\

&
&
& 0.60
& 81--352 \\

\hline

EXO-200
& $>1.6\times10^{25}$
& 677
& 1.27
& 88--382 \\

&
&
& 1.00
& 142--616 \\

&
&
& 0.80
& 222--962 \\

&
&
& 0.60
& 395--1711 \\

\hline

NEXT-100
& $\sim2.85\times10^{25}$
& 507
& 1.27
& 66--287 \\

&
&
& 1.00
& 106--461 \\

&
&
& 0.80
& 166--720 \\

&
&
& 0.60
& 295--1280 \\

\hline

LZ
& $\sim1.06\times10^{26}$
& 263
& 1.27
& 34--149 \\

&
&
& 1.00
& 55--239 \\

&
&
& 0.80
& 86--374 \\

&
&
& 0.60
& 153--664 \\

\hline

nEXO
& $\sim1.35\times10^{28}$
& 7.37
& 1.27
& 0.96--4.16 \\

&
&
& 1.00
& 1.54--6.70 \\

&
&
& 0.80
& 2.40--10.47 \\

&
&
& 0.60
& 4.27--18.63 \\

\end{tabular}
\end{ruledtabular}
\end{table*}
The results demonstrate that the inferred Majorana mass sensitivity varies substantially depending on the adopted NME calculation and the assumed degree of axial-current quenching.  The spread originating from theoretical uncertainties is comparable to, and in some cases larger than, the experimental uncertainty itself.
Current experiments such as KamLAND-Zen, EXO-200, NEXT-100, and LZ primarily probe the inverted-ordering region of neutrino masses, corresponding roughly to
\[\langle m_{\beta\beta}\rangle_{\rm IH}\sim 15\text{--}50~{\rm meV}.\]
In contrast, the projected nEXO sensitivity approaches the
normal-ordering regime,
\[
\langle m_{\beta\beta}\rangle_{\rm NH}\sim 1\text{--}10~{\rm meV},
\]
particularly in the optimistic scenario of large NMEs and weak quenching. However, strong quenching of $g_A$ significantly weakens the achievable mass reach even for ton-scale detectors.

These results highlight that future progress in neutrinoless double beta decay searches will require not only larger isotope exposures and lower background levels, but also improved understanding of nuclear structure effects, axial-current renormalization, and theoretical uncertainties in NME calculations.
\subsection*{Conclusion on Majorana Mass Sensitivity}

From the comparison of KamLAND-Zen, EXO-200, and nEXO, it is evident that the sensitivity to the effective Majorana neutrino mass improves significantly with increased exposure and reduced background, as seen in next-generation experiments such as nEXO. 

However, the extracted mass limits are strongly affected by theoretical uncertainties, particularly the choice of nuclear matrix element and the axial-vector coupling \(g_A\). For all experiments, the allowed range of \(\langle m_{\beta\beta}\rangle\) broadens substantially under quenching of \(g_A\), with the mass limits weakening by a factor of \(\sim 4\text{--}5\) when going from the unquenched value \(g_A=1.27\) to strongly quenched scenarios \(g_A\approx0.6\).

Thus, while experimental advancements push the half-life sensitivity to unprecedented levels, the precise determination of the Majorana mass remains limited by nuclear theory uncertainties. A consistent interpretation of future results will therefore require improved constraints on nuclear matrix elements and a better understanding of axial coupling quenching in nuclei.

\subsection{Propagation of Theoretical and Experimental Uncertainties}
\begin{figure}[H]
 \centering
\includegraphics[width=1.0\linewidth]{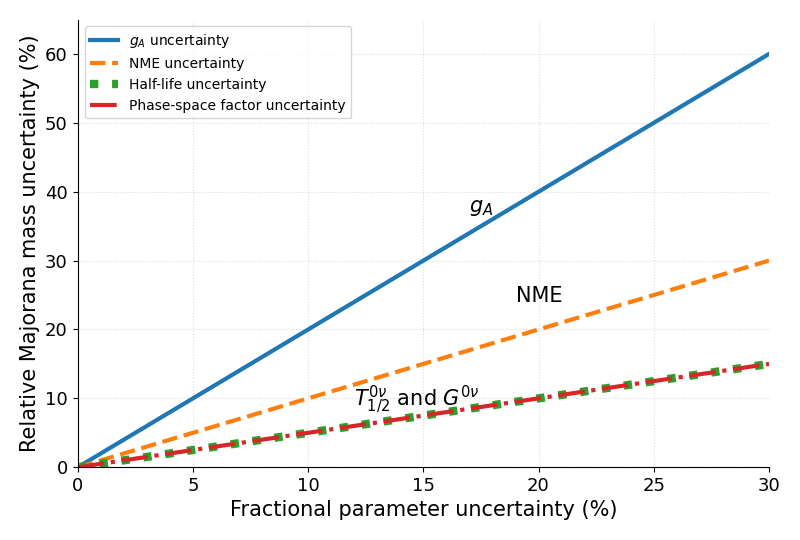}
\caption{Propagation of theoretical and experimental uncertainties in the effective Majorana mass sensitivity. The relative uncertainty associated with the axial-vector coupling constant \(g_A\) is amplified by a factor of two due to the dependence \(m_{\beta\beta}\propto g_A^{-2}\), while the nuclear matrix element (NME) contribution enters linearly. In contrast, uncertainties associated with the half-life \(T_{1/2}^{0\nu}\) and phase-space factor \(G^{0\nu}\) are suppressed by a factor of one half, reflecting their weaker dependence on the extracted Majorana mass sensitivity.}
\label{fig:Uncertaniries}
\end{figure}

In  Eq.~(\ref{eq:maj_mass_xe}), for compactness, let us assume
\[m \equiv \langle m_{\beta\beta}\rangle .\]
Now the relevant partial derivatives  of this equation are
\begin{align}
\frac{\partial m}{\partial g_A}&=-2\frac{m}{g_A},
\\
\frac{\partial m}{\partial M^{0\nu}}&=-\frac{m}{M^{0\nu}},
\\
\frac{\partial m}{\partial T_{1/2}^{0\nu}}&=-\frac{1}{2}\frac{m}{T_{1/2}^{0\nu}},
\\
\frac{\partial m}{\partial G^{0\nu}}&=-\frac{1}{2} \frac{m}{G^{0\nu}}.
\end{align}
Therefore, assuming independent uncertainties, the relative uncertainty is
\begin{equation}
\begin{aligned}
\left(\frac{\delta m}{m}
\right)^2=4\left(\frac{\delta g_A}{g_A}\right)^2+
\left(\frac{\delta M^{0\nu}}{M^{0\nu}}\right)^2
+\frac{1}{4}\left(
\frac{\delta T_{1/2}^{0\nu}}{T_{1/2}^{0\nu}}
\right)^2+ \\
\frac{1}{4}\left(\frac{\delta G^{0\nu}}{G^{0\nu}}
\right)^2 .
\end{aligned}
\end{equation}
This expression shows that the uncertainty in $g_A$ is amplified by a factor of two in the relative mass uncertainty because
$\langle m_{\beta\beta}\rangle \propto g_A^{-2}$.
The uncertainty in the NME enters linearly, since
$\langle m_{\beta\beta}\rangle \propto |M^{0\nu}|^{-1}$.
By contrast, uncertainties in the half-life and phase-space factor are suppressed by a factor of one half because
$\langle m_{\beta\beta}\rangle \propto (T_{1/2}^{0\nu})^{-1/2}$ and
$\langle m_{\beta\beta}\rangle \propto (G^{0\nu})^{-1/2}$.
Thus, the dominant theoretical limitations arise from the axial coupling constant and the nuclear matrix element. This explains why different NME models and assumptions about $g_A$ quenching can produce large variations in the inferred Majorana mass even for the same experimental half-life sensitivity. The impact of these uncertainties is illustrated in Figure~\ref{fig:Uncertaniries}, where the propagation of both theoretical and experimental contributions to the effective Majorana mass sensitivity is shown. The uncertainty associated with the axial-vector coupling constant \(g_A\) is particularly significant because the effective Majorana mass scales as \(m_{\beta\beta}\propto g_A^{-2}\), leading to an amplification of its relative uncertainty by a factor of two. The uncertainty arising from the nuclear matrix element enters linearly and therefore also contributes substantially to the overall spread in the extracted mass range. By contrast, the uncertainties associated with the half-life \(T_{1/2}^{0\nu}\) and the phase-space factor \(G^{0\nu}\) are comparatively suppressed, since they affect the Majorana mass only through a square-root dependence. Consequently, even when an experiment achieves a well-defined half-life sensitivity, the inferred range of \(m_{\beta\beta}\) remains broadened due to nuclear-structure modeling and axial-coupling ambiguities.
This demonstrates that even for a fixed experimental half-life limit, the extracted Majorana mass sensitivity spans a finite interval due to nuclear-structure and axial-coupling uncertainties.
\section{Estimation of Effective Majorana Mass from Neutrino Oscillation Parameters}

In the previous section, the effective Majorana neutrino mass was inferred from neutrinoless double beta decay half-life sensitivities. In the present section, we estimate the allowed range of $\langle m_{\beta\beta}\rangle$ directly from neutrino oscillation parameters within the standard three-flavor framework. Unlike the half-life-based extraction, this approach is independent of detector response, isotope exposure, and nuclear matrix element calculations, and therefore provides the theoretically
allowed parameter space that neutrinoless double beta decay experiments attempt to probe.
For three Majorana neutrinos, the effective Majorana mass is given by
\begin{equation}
\langle m_{\beta\beta} \rangle =
\left|m_1 U_{e1}^2+m_2 U_{e2}^2 e^{i\alpha_{21}}+m_3 U_{e3}^2e^{i\alpha_{31}}
\right|,
\end{equation}
where $m_i$ denote the eigenvalues of the neutrino mass, $U_{ei}$ are the elements of the PMNS lepton mixing matrix, and $\alpha_{21}$ and $\alpha_{31}$ are the unknown Majorana CP-violating phases.
Using the standard PMNS parameterization, the expression above becomes
\begin{equation}
\langle m_{\beta\beta}\rangle
=\left|c_{12}^2 c_{13}^2\, m_1+
s_{12}^2 c_{13}^2\, m_2\, e^{i\alpha_{21}}
+s_{13}^2\, m_3\, e^{i\alpha_{31}}\right|,
\label{eq:majorana_pmns}
\end{equation}
where
\[c_{ij} \equiv \cos\theta_{ij},\qquad_{ij} \equiv \sin\theta_{ij},\]
and the Majorana phases satisfy
\begin{equation}
0 \le \alpha_{21},\,\alpha_{31} < 2\pi.
\end{equation}
The allowed range of $\langle m_{\beta\beta}\rangle$
is determined using the experimentally measured neutrino mixing angles and mass-squared differences. For normal hierarchy (NH),
\begin{align}
m_2 &= \sqrt{m_1^2 + \Delta m_{21}^2},
\\
m_3 &= \sqrt{m_1^2 + \Delta m_{31}^2},
\end{align}
while for the inverted-hierarchy (IH),
\begin{align}
m_1 &= \sqrt{m_3^2 + |\Delta m_{31}^2|},
\\
m_2 &= \sqrt{m_3^2 + |\Delta m_{31}^2| + \Delta m_{21}^2}.
\end{align}
The resulting parameter space strongly depends on the unknown Majorana phases. Constructive interference between the contributing terms enhances $\langle m_{\beta\beta}\rangle$, whereas destructive interference can significantly suppress the effective mass, particularly in the normal ordering scenario.
The determination of $\langle m_{\beta\beta}\rangle$
therefore depends sensitively on the neutrino oscillation mixing angles; $\theta_{12}$, $\theta_{13}$, and mass-squared difference; $\Delta m_{21}^{2}$, and $\Delta m_{31}^{2}$.
Any uncertainty in these quantities propagates directly into the inferred Majorana mass region.
In neutrino oscillation experiments, the extraction of oscillation
parameters is affected by FSI, which modify the observable final-state particle distributions through scattering, absorption, and charge-exchange processes inside the nuclear medium. These effects distort the reconstructed neutrino energy spectrum and introduce systematic uncertainties in the measured oscillation parameters. Since the effective Majorana mass is directly connected to these parameters through Eq.~\eqref{eq:majorana_pmns}, FSI-induced uncertainties propagate indirectly into the estimation of $\langle m_{\beta\beta}\rangle$ for both normal hierarchy and inverted hierarchy neutrino mass scenarios.
Unlike the half-life-based extraction discussed earlier, the present
oscillation-based framework is independent of nuclear matrix element
uncertainties and detector-specific systematic effects. Consequently,
this approach defines the theoretically allowed region of
$\langle m_{\beta\beta}\rangle$ that future neutrinoless double beta decay experiments must probe in order to test the Majorana nature of neutrinos and distinguish between the normal and inverted neutrino mass orderings.
\subsection{Best-Fit Oscillation Parameters}
Estimation of the effective Majorana neutrino mass is performed using the global best-fit oscillation parameters corresponding to the IC19 analysis without Super-Kamiokande atmospheric data \cite{Esteban:2024eli}. The adopted values are
\begin{equation}
\sin^2\theta_{12}=0.307,
\qquad
\sin^2\theta_{13}=
\begin{cases}
0.02195 & (\mathrm{NH}),
\\
0.02215 & (\mathrm{IH}),
\end{cases}
\end{equation}

\begin{equation}
\Delta m_{21}^2=7.49\times10^{-5}\ {\rm eV}^2,
\end{equation}

\begin{equation}
\Delta m_{3\ell}^2=\begin{cases}
+2.534\times10^{-3}\ {\rm eV}^2
& (\mathrm{NH}),
\\
-2.510\times10^{-3}\ {\rm eV}^2
& (\mathrm{IH}).
\end{cases}
\end{equation}
These oscillation parameters determine the neutrino mass spectrum and therefore directly constrain the allowed range of the effective Majorana mass.
\subsubsection{Normal Hierarchy}
For normal ordering (NH), the neutrino mass eigenvalues are expressed as
\begin{align}
m_1 &= m_{\rm lightest},
\\
m_2 &= \sqrt{
m_1^2 + \Delta m_{21}^2
},
\\
m_3 &= \sqrt{
m_1^2 + \Delta m_{3\ell}^2
}.
\end{align}
In the hierarchical limit,$m_1 \rightarrow 0$,the neutrino spectrum is dominated by the measured mass splittings, giving
\begin{align}
m_2 &\approx 8.65~{\rm meV},
\\
m_3 &\approx 50.34~{\rm meV}.
\end{align}
Under this approximation, the contribution from $m_1$ becomes negligible, and Eq.~\eqref{eq:majorana_pmns} reduces to
\begin{equation}
m_{\beta\beta}^{\rm NH}=
\left|s_{12}^{2}c_{13}^{2}m_{2}e^{i\alpha_{21}}
+s_{13}^{2}m_{3}e^{i\alpha_{31}}\right|.
\end{equation}
The effective Majorana mass is therefore determined by the interference between these two terms through the unknown Majorana phases $\alpha_{21}$ and $\alpha_{31}$.
Constructive interference produces the maximum effective mass, while
destructive interference suppresses the observable signal.
Considering the extreme interference limits here and substituting the best-fit values of oscillation parameters, one obtains
the allowed range in the normal-hierarchy and the limit becomes
\begin{equation}
m_{\beta\beta}^{\rm NH}
\approx
1.5\text{--}3.7~{\rm meV}.
\end{equation}
This result demonstrates one of the characteristic features of the normal hierarchy that is the possibility of strong cancellation between the individual contributions due to the unknown Majorana phases.
Consequently, the effective Majorana mass can remain extremely small even when neutrinos possess non-zero masses. This places the normal-ordering region well below the sensitivity reach of most present-generation neutrinoless double beta decay experiments.
The predicted normal-hierarchy band is sensitive to small variations in the oscillation parameters, particularly $\theta_{12}$ and $\theta_{13}$, because the effective Majorana mass in this regime arises from a delicate interference between the $m_2$ and $m_3$ contributions. Consequently, systematic effects that modify the precision extraction of oscillation
parameters can slightly shift the allowed $m_{\beta\beta}^{\rm NH}$ region. Since Final State Interactions (FSI) contribute to the systematic uncertainty budget in oscillation analyses,
their impact becomes increasingly relevant when probing the meV-scale sensitivity region associated with the normal mass ordering.
FSI-induced systematic uncertainties in neutrino oscillation measurements can propagate into the extracted effective Majorana mass through the oscillation parameters entering Eq.~\eqref{eq:majorana_pmns}. The resulting
uncertainty is generally subdominant compared with nuclear matrix element and axial-coupling uncertainties, but may become relevant in the normal-hierarchy  regime where $m_{\beta\beta}\sim\mathcal{O}(1~{\rm meV})$ and strong phase cancellations occur.
\subsubsection{Inverted Hierarchy}
For inverted hierarchy (IH), the neutrino mass eigenvalues are given by
\begin{align}
m_3 &= m_{\rm lightest},
\\
m_1 &= \sqrt{
m_3^{2}
+
|\Delta m_{3\ell}^{2}|
},
\\
m_2 &= \sqrt{
m_3^2
+
|\Delta m_{3\ell}^{2}|
+
\Delta m_{21}^{2}
}.
\end{align}
In the hierarchical limit, $m_3 \rightarrow 0$,
the neutrino spectrum is dominated by the atmospheric mass splitting, leading to
\begin{align}
m_1 &\approx 50.10~{\rm meV},
\\
m_2 &\approx 50.85~{\rm meV}.
\end{align}
Under this approximation, the contribution from $m_3$ becomes negligible, and Eq.~\eqref{eq:majorana_pmns} reduces to
\begin{equation}
m_{\beta\beta}^{\rm IH}
=
\left|
c_{12}^{2}c_{13}^{2}m_{1}
+
s_{12}^{2}c_{13}^{2}m_{2}e^{i\alpha_{21}}
\right|.
\end{equation}
The effective Majorana mass is controlled primarily by the interference between these two nearly degenerate contributions through the unknown Majorana phase $\alpha_{21}$. Unlike the normal hierarchy case, complete cancellation is not possible because the magnitudes of the two terms remain comparatively large.
Considering the limiting cases of constructive and destructive
interference, and substituting the best fit values of oscillation parameters one obtains he allowed range in the Inverted-hierarchy and the effective Majorana mass limit  in this hierarchy becomes
\begin{equation}
m_{\beta\beta}^{\rm IH}
\approx
18.7\text{--}49.2~{\rm meV}.
\end{equation}
A characteristic feature of the inverted hierarchy is the persistence of a substantial lower bound on
$m_{\beta\beta}$.
Since the dominant contributions arise from two nearly degenerate mass eigenstates, destructive interference cannot completely suppress the effective mass. Consequently, the inverted hierarchy region remains within the sensitivity reach of several present and next-generation neutrinoless double beta decay experiments.
In this regime, the influence of FSI-induced uncertainties on the predicted effective mass band is relatively moderate because the dominant contribution originates from the large atmospheric mass scale. Therefore, small shifts in the extracted oscillation parameters produce only limited variations in
the predicted $m_{\beta\beta}^{\rm IH}$ range compared with the normal hierarchy scenario.
\subsubsection{Final Results and Physical Implications}
Using the best-fit oscillation parameters in the hierarchical limit, the effective Majorana mass is obtained as
\begin{equation}
m_{\beta\beta}^{\rm NH}
\approx
1.5\text{--}3.7~{\rm meV},
\qquad
m_{\beta\beta}^{\rm IH}
\approx
18.7\text{--}49.2~{\rm meV}.
\end{equation}
These results reveal a strong separation between the normal hierarchy and inverted hierarchy regions. In the normal hierarchy, the effective Majorana mass is strongly suppressed due to the smallness of the lightest neutrino mass and the possibility of destructive interference among the contributing terms through the unknown Majorana phases. Consequently, the predicted mass range lies at the few-meV level.
In contrast, the inverted hierarchy is characterized by two nearly
degenerate mass eigenstates that contribute coherently to the effective Majorana mass. Since complete cancellation between the dominant terms is not possible, a substantial non-zero lower bound on $m_{\beta\beta}$ persists even for unfavorable phase configurations. This makes the inverted hierarchy region significantly more accessible to present and future neutrinoless double beta decay experiments.
These hierarchy-dependent features have important experimental
implications. Sensitivities at the level of $\sim10\text{--}20~{\rm meV}$ are sufficient to probe a major fraction of the inverted hierarchy region, whereas testing the normal hierarchy requires sensitivity in the sub-10 meV regime. Reaching such precision remains one of the principal challenges for next-generation ton-scale detectors.
As experiments approach the meV-scale sensitivity frontier, systematic effects associated with event reconstruction and detector response become increasingly important. Reconstruction uncertainties analogous to FSI, including scattering, energy loss, bremsstrahlung, and topology-changing processes inside the detector medium, can distort the reconstructed energy spectrum and broaden the region of interest. Such effects propagate into the extracted half-life sensitivity and consequently into the inferred Majorana mass.
\begin{figure}
    \centering
    \includegraphics[width=\linewidth]{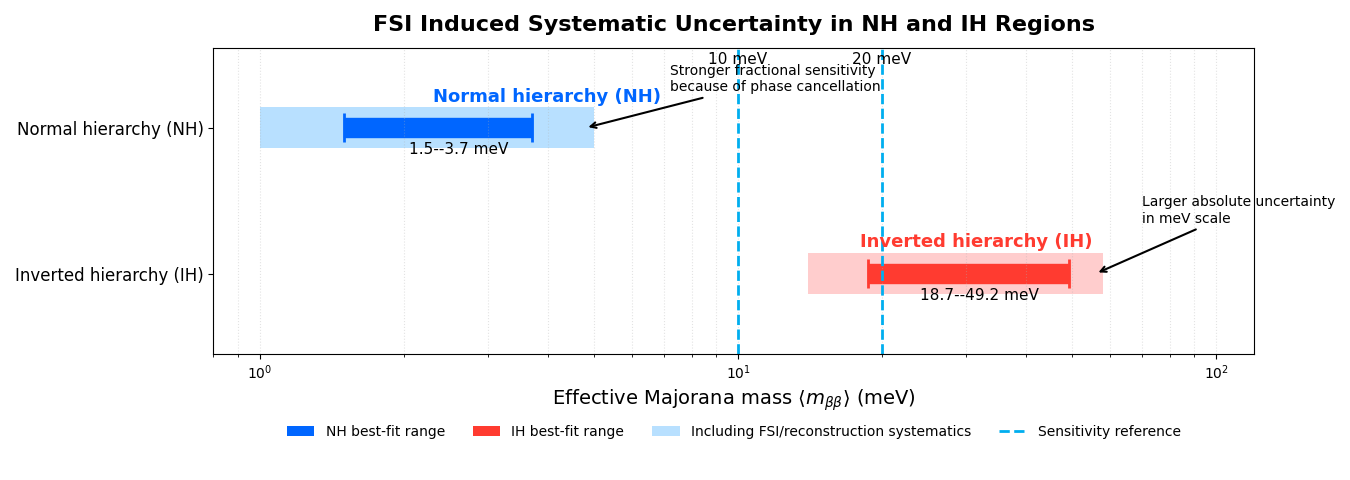}
    \caption{FSI induced systematic uncertainty in the effective Majorana mass for Normal and Inverted Hierarchies.}
    \label{fig:fsi_uncertanity}
\end{figure}
The Fig.~\ref{fig:fsi_uncertanity} shows the effect of FSI/reconstruction-induced uncertainties on the effective Majorana mass 
$\langle m_{\beta\beta} \rangle$ for both the NH and IH regions. 
The horizontal axis represents the effective Majorana mass in meV on a logarithmic scale.
The dark blue horizontal bands denote the best-fit allowed ranges obtained from oscillation parameters, while the surrounding light blue shaded regions represent the additional spread introduced by 
FSI and reconstruction-related systematic uncertainties. 
For the NH case, the allowed range lies approximately between 
$1.5$--$3.7$ meV, whereas for the IH case the allowed interval extends from about $18.7$--$49.2$ meV.

The blue dashed vertical lines at 10 meV and 20 meV indicate representative experimental sensitivity benchmarks for next-generation neutrinoless double beta decay experiments. 
The NH region appears more sensitive to fractional uncertainties because the effective mass itself is strongly suppressed 
due to destructive Majorana phase cancellation. 
Consequently, even a small absolute uncertainty produces a comparatively large fractional variation in the NH region.

In contrast, the IH region exhibits a larger absolute uncertainty in terms of meV scale, as reflected by the broader 
light-blue systematic band. However, because the effective mass values are substantially larger in IH, 
the relative (fractional) impact of these uncertainties is comparatively moderate. 
This demonstrates that future experiments require significantly improved reconstruction precision and background control 
to probe the NH regime reliably. For representative reconstruction uncertainties of order $5\%\text{--}15\%$, the induced uncertainty in the extracted effective Majorana mass can reach approximately $10\%\text{--}30\%$.
While this effect remains subdominant compared with uncertainties arising from nuclear matrix elements and axial-coupling quenching, it becomes particularly relevant in the normal hierarchy region where the expected signal itself is only at the few-meV level.

These results indicate that future improvements in neutrinoless double beta decay sensitivity cannot rely solely on increased isotope exposure. Next-generation experiments must simultaneously achieve ultra-low background levels, improved detector response modeling, precise energy reconstruction, and reduced theoretical uncertainties in nuclear matrix elements in order to fully probe the normal hierarchy parameter space.
\subsection{Hierarchy Reach of Xenon-Based Experiments and FSI Implications}
Using the effective Majorana mass ranges derived from neutrino oscillation parameters, we now discuss the hierarchy reach of current and next-generation \(^{136}\)Xe-based neutrinoless double beta decay experiments in the presence of FSI and reconstruction-related uncertainties.
\begin{itemize}
\item[(i)] \textbf{KamLAND-Zen:}
\[\langle m_{\beta\beta}\rangle < 28\text{--}122~\mathrm{meV}.\]
The KamLAND-Zen sensitivity overlaps significantly with the inverted-hierarchy (IH) region, whose characteristic range lies approximately between 
\(\sim 19\text{--}50~\mathrm{meV}\). 
Its sensitivity extends through the upper and intermediate IH band, while also constraining a substantial fraction of the quasi-degenerate regime. 
However, the lower edge of the IH parameter space remains beyond its reach. 
Consequently, although KamLAND-Zen places important limits on the effective Majorana mass, a null observation cannot completely exclude the inverted hierarchy. 
FSI and reconstruction uncertainties further broaden the experimentally accessible region, particularly near the lower IH boundary where precise energy reconstruction becomes increasingly important.
\item[(ii)] \textbf{LZ:}
\[\langle m_{\beta\beta}\rangle < 53\text{--}164~\mathrm{meV}.\]
The LZ sensitivity primarily probes the quasi-degenerate region and overlaps only with the upper portion of the IH band. 
Its reach does not extend to the characteristic IH floor, and therefore a large fraction of the inverted-ordering parameter space remains inaccessible. 
In this regime, FSI-related effects mainly contribute to absolute uncertainties in the extracted mass scale, although the overall hierarchy sensitivity remains limited by the comparatively high threshold.
\item[(iii)] \textbf{NEXT-100:}
\[\langle m_{\beta\beta}\rangle < 80\text{--}160~\mathrm{meV}.\]
The sensitivity of NEXT-100 lies predominantly in the quasi-degenerate region, where the three neutrino mass eigenstates become nearly degenerate. 
Its overlap with the IH region is restricted to the extreme upper edge, resulting in only marginal sensitivity to the inverted hierarchy. 
At these larger effective masses, reconstruction and FSI uncertainties contribute mainly to the absolute uncertainty, while their fractional impact remains comparatively moderate.
\item[(iv)] \textbf{nEXO:}
\[\langle m_{\beta\beta}\rangle < 4.7\text{--}20.3~\mathrm{meV}.\]
The projected sensitivity of nEXO represents a major advancement in hierarchy exploration. 
Its reach fully covers the inverted-hierarchy band and extends down to its lower boundary, providing the capability to either observe a \(0\nu\beta\beta\) signal or strongly exclude the IH scenario within the standard light Majorana neutrino exchange framework. 
Furthermore, the lower sensitivity range of nEXO begins to approach the normal-hierarchy (NH) region, where 
\[\langle m_{\beta\beta}\rangle_{\rm NH} \sim 1.5\text{--}3.7~\mathrm{meV}.\]
However, the NH regime remains experimentally challenging because strong Majorana phase cancellations suppress the effective mass to the few-meV scale. 
In this region, even small FSI/reconstruction effects generate large fractional uncertainties, requiring extremely precise detector calibration, energy resolution, and background suppression.
\end{itemize}
From the estimated sensitivities of the above experiments, it can be concluded that present-generation detectors such as KamLAND-Zen, LZ, and NEXT-100 primarily constrain the quasi-degenerate and upper inverted-hierarchy regions, but they do not achieve sensitivity to the generic normal-hierarchy parameter space. 
Their hierarchy reach is additionally affected by FSI and reconstruction-related systematic uncertainties, which broaden the allowed effective-mass intervals. 
In contrast, next-generation experiments such as nEXO are designed to achieve the sensitivity necessary for a decisive exploration of the inverted hierarchy and to approach the experimentally demanding normal-hierarchy regime.
\subsubsection{Exposure Requirement for $^{136}Xe$ at $3\sigma$ and $5\sigma$ Confidence Levels}
Using the effective Majorana mass sensitivities obtained in Section~\ref{sec:Experiment}, we estimate the detector exposure required to probe representative regions of the IH and NH parameter space. In the background-dominated regime, the effective Majorana mass sensitivity scales with detector exposure as
\begin{equation}
m_{\beta\beta}^{\rm sens}\propto \Sigma^{-1/4},
\end{equation}
which is already expressed in Equation~\ref{eq:majorna-mass}
\\

Using the estimated Majorana neutrino mass sensitivities at $3\sigma$ and $5\sigma$ together with the corresponding exposures

\begin{align}
\Sigma_0^{\rm LZ} &= 0.236~{\rm ton\cdot yr}, \qquad
\Sigma_0^{\rm NEXT} = 0.275~{\rm ton\cdot yr}, \nonumber \\
\Sigma_0^{\rm KLZ} &= 1.13~{\rm ton\cdot yr}, \qquad                                    
\Sigma_0^{\rm nEXO} = 0.10~{\rm ton\cdot yr},
\end{align}

we evaluate the exposure required to probe representative IH and NH regions.

For the inverted-ordering region, we consider

\[
m_{\beta\beta}=49.2~{\rm meV}
\quad \text{and} \quad
18.7~{\rm meV},
\]

corresponding approximately to the upper and lower boundaries of the IH band.

For the NH region, we consider

\[
m_{\beta\beta}=3.7~{\rm meV}
\quad \text{and} \quad
1.4~{\rm meV},
\]

representing the upper and lower portions of the NH parameter space.

The exposure estimates obtained for IH and NH regions are summarized in Tables~\ref{tab:exposure_requirements_IH} and~\ref{tab:exposure_requirements_NH}. 
The results clearly demonstrate the extremely strong scaling behavior 
\(
\Sigma_{\rm req}\propto m_{\beta\beta}^{-4},
\)
which leads to a rapid increase in the required exposure as the target effective Majorana mass decreases.

\begin{figure}
    \centering
    \includegraphics[width=\linewidth]{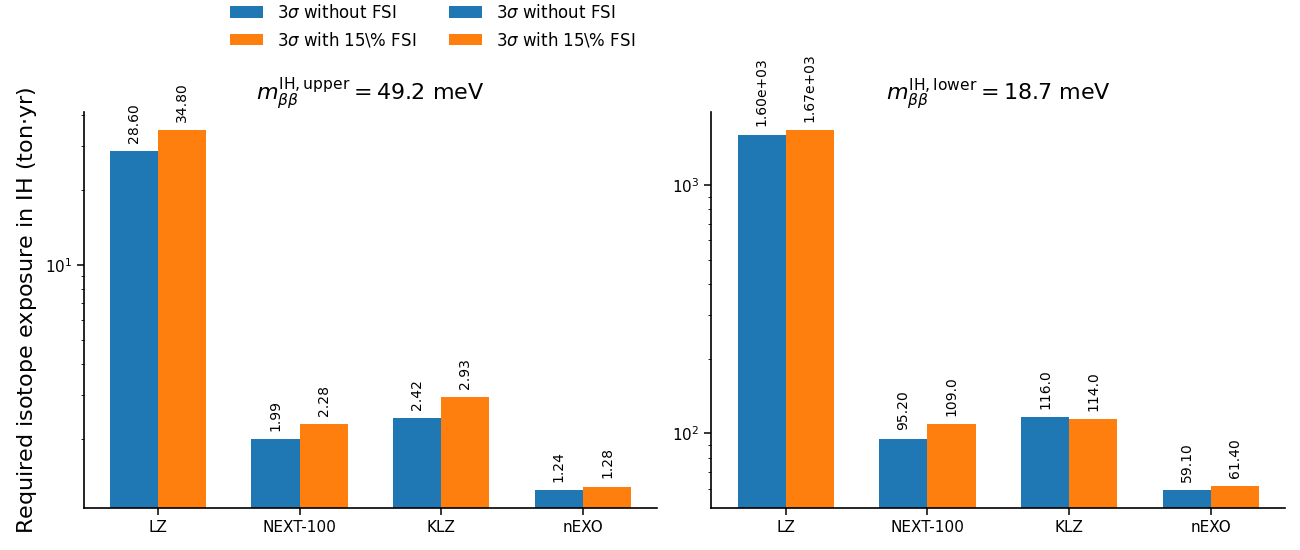}
   \caption{Required isotope exposure as a function of effective Majorana mass for different \(^{136}\)Xe-based experiments in IH region at \(3\sigma\) confidence level.}
    \label{fig:IH}
\end{figure}

The corresponding exposure scaling in the IH region is illustrated in Fig.~\ref{fig:IH}. 
The figure shows that the required exposure increases steeply toward the lower edge of the IH band near 
\(m_{\beta\beta}\sim18.7~\mathrm{meV}\), reflecting the quartic dependence of exposure on the Majorana mass sensitivity.

\begin{figure}
    \centering
    \includegraphics[width=\linewidth]{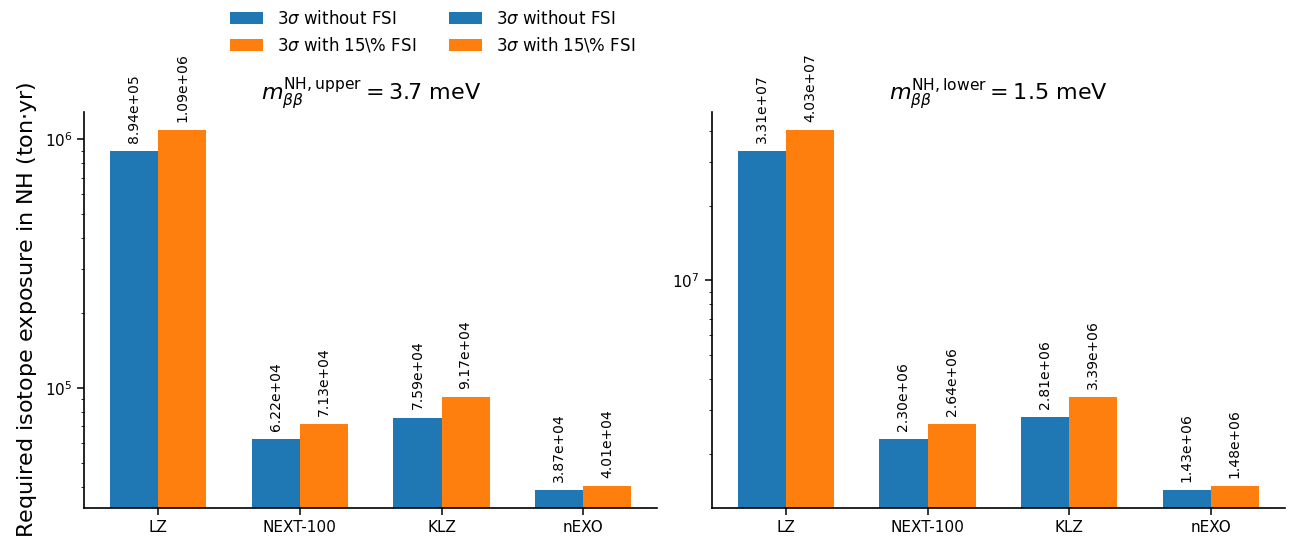}
   \caption{Required isotope exposure as a function of effective Majorana mass for different \(^{136}\)Xe-based experiments in the NH region at \(3\sigma\) confidence level.}
    \label{fig:nh}
\end{figure}

The NH exposure scaling is shown in Fig.~\ref{fig:nh}. In comparison with the IH region, the NH region requires several orders of magnitude larger exposure due to the much smaller effective Majorana mass scale. The effect of FSI-induced degradation is also visible, producing a moderate upward shift in the exposure requirement for all experiments.
\begin{table}[htbp]
\centering
 \caption{Required isotope exposures for different \(^{136}\)Xe-based experiments to probe the upper and lower inverted-hierarchy regions at \(3\sigma\) confidence level, with and without \(15\%\) FSI effects.
}
\label{tab:exposure_requirements_IH}
\small
\setlength{\tabcolsep}{2.0pt}
\renewcommand{\arraystretch}{1.0}
\begin{tabular}{lcccc}
\hline
\hline
\begin{tabular}{l}
Exposure Req. \\
(ton\(\cdot\)yr)
\end{tabular}
& LZ & NEXT-100 & KLZ & nEXO \\
\hline
\multicolumn{5}{c}{\(m_{\beta\beta}=49.2~\mathrm{meV}\)} \\
\hline
\(3\sigma\)
& \(2.86\times10^{1}\)
& \(1.99\)
& \(2.42\)
& \(1.24\)
\\

\(3\sigma+\)FSI
& \(3.48\times10^{1}\)
& 2.28
& 2.93
& 1.28
\\

\(5\sigma\)
& \(9.46\times10^{1}\)
& \(7.78\)
& \(8.15\)
& \(5.85\)
\\

\(5\sigma+\)FSI
& \(1.09\times10^{2}\)
&  8.49
& 10.50
& 6.22
\\

\hline

\multicolumn{5}{c}{\(m_{\beta\beta}=18.7~\mathrm{meV}\)} \\
\hline

\(3\sigma\)
& \(1.6\times10^{3}\)
& \(9.52\times10^{1}\)
& \(1.16\times10^{2}\)
& \(5.91\times10^{1}\)
\\

\(3\sigma+\)FSI
& \(1.67\times10^{3}\)
& \(1.09\times10^{2}\)
& \(1.14\times10^{2}\)
& \(6.14\times10^{1}\)
\\

\(5\sigma\)
& \(4.28\times10^{3}\)
& \(3.73\times10^{2}\)
& \(3.91\times10^{2}\)
& \(2.81\times10^{2}\)
\\

\(5\sigma+\)FSI
& \(5.27\times10^{3}\)
& \(4.07\times10^{2}\)
& \(5.03\times10^{2}\)
&\(2.98\times10^{2}\)
\\

\hline
\hline

\end{tabular}

\end{table}

\begin{table}[htbp]
\centering

\caption{
Required isotope exposures for different \(^{136}\)Xe-based experiments to probe the upper and lower normal-hierarchy regions at \(3\sigma\) confidence level, with and without \(15\%\) FSI effects.}

\label{tab:exposure_requirements_NH}

\small
\setlength{\tabcolsep}{2.0pt}
\renewcommand{\arraystretch}{1.0}

\begin{tabular}{lcccc}

\hline
\hline

\begin{tabular}{l}
Exposure Req. \\
(ton\(\cdot\)yr)
\end{tabular}
& LZ & NEXT-100 & KLZ & nEXO \\
\hline

\multicolumn{5}{c}{\(m_{\beta\beta}=3.7~\mathrm{meV}\)} \\
\hline

\(3\sigma\)
& \(8.94\times10^{5}\)
& \(6.22\times10^{4}\)
& \(7.59\times10^{4}\)
& \(3.87\times10^{4}\)
\\

\(3\sigma+\)FSI
& \(1.09\times10^{6}\)
& \(7.13\times10^{4}\)
& \(9.17\times10^{4}\)
& \(4.01\times10^{4}\)
\\

\(5\sigma\)
& \(2.83\times10^{6}\)
& \(2.43\times10^{5}\)
& \(2.55\times10^{5}\)
& \(1.83\times10^{5}\)
\\

\(5\sigma+\)FSI
& \(3.43\times10^{6}\)
& \(2.43\times10^{5}\)
& \(3.28\times10^{5}\)
& \(1.95\times10^{5}\)
\\

\hline

\multicolumn{5}{c}{\(m_{\beta\beta}=1.5~\mathrm{meV}\)} \\
\hline

\(3\sigma\)
& \(3.31\times10^{7}\)
& \(2.30\times10^{6}\)
& \(2.81\times10^{6}\)
& \(1.43\times10^{6}\)
\\

\(3\sigma+\)FSI
& \(4.03\times10^{7}\)
& \(2.64\times10^{6}\)
& \(3.39\times10^{6}\)
& \(1.48\times10^{6}\)
\\

\(5\sigma\)
& \(1.05\times10^{8}\)
& \(9.01\times10^{6}\)
& \(9.44\times10^{6}\)
& \(6.78\times10^{6}\)
\\

\(5\sigma+\)FSI
& \(1.27\times10^{8}\)
&\(9.83\times10^{6}\)
&\(1.21\times10^{7}\)
& \(7.2\times10^{6}\)
\\
\hline
\hline
\end{tabular}
\end{table}
The calculations show that ton-scale next-generation experiments such as nEXO can probe a substantial fraction of the IH region. However, exploring the NH regime requires exposures several orders of magnitude larger because of the strong suppression of the effective Majorana mass.
Therefore, future experiments aiming to probe the NH region will require not only very large exposures, but also substantial improvements in background reduction, detector resolution, and systematic control.
\section{Discussion}
\label{sec:discussion}
The observation of neutrinoless double beta decay (\(0\nu\beta\beta\)) would establish lepton-number violation and confirm the Majorana nature of neutrinos. In addition, it would provide direct information on the absolute neutrino mass scale through the effective Majorana mass parameter,
\[\langle m_{\beta\beta} \rangle .\]
A meaningful interpretation of present and future \(^{136}\)Xe-based experiments therefore requires a statistically consistent framework that simultaneously incorporates isotope exposure, detector resolution, background fluctuations, NME uncertainties,d reconstruction-related systematic effects.
In the present work, we developed a unified exposure-based framework using the relation
\[N_b = \mathrm{BI}\,\Delta E\,\Sigma_{\mathrm{iso}},\]
where \(N_b\) denotes the expected background counts in the region of interest (ROI), \(\mathrm{BI}\) is the background index, \(\Delta E\) is the ROI width, and \(\Sigma_{\mathrm{iso}}\) represents the isotope exposure. This formulation allows different detector technologies to be compared on an equivalent statistical basis.
For a common matched exposure of
\[\Sigma_{\mathrm{iso}} = 100~\mathrm{kg\cdot yr},\]
the estimated background expectations are
\begin{align*}
N_b^{\mathrm{LZ}} &\approx 15.10,
\qquad
N_b^{\mathrm{NEXT\text{-}100}} \approx 1.16,
\\[4pt]
N_b^{\mathrm{KLZ}} &\approx 5.39,
\qquad
N_b^{\mathrm{nEXO}} \approx 0.059.
\end{align*}
These estimated background number is the interplay between energy resolution, ROI definition, and detector-specific background rejection capability. Experiments nEXO and NEXT-100 benefits from superior topological discrimination and a narrow ROI, resulting in the smallest expected background contribution. KamLAND-Zen, despite its large exposure capability, operates with a broader ROI characteristic of liquid scintillator detectors. The LZ estimate, derived from a detector optimized primarily for dark-matter searches, exhibits comparatively larger effective background contributions in the \(0\nu\beta\beta\) ROI.

The statistical analysis also highlights the transition between background-free and background-dominated regimes. In the background-free limit (\(N_b \ll 1\)),
\[T_{1/2}^{0\nu} \propto \Sigma_{\mathrm{iso}},\]
whereas in the background-limited regime (\(N_b \gg 1\)),
\[T_{1/2}^{0\nu} \propto \sqrt{\Sigma_{\mathrm{iso}}}.\]
This transition imposes a fundamental limitation on sensitivity improvement and demonstrates that future experimental progress cannot rely exclusively on increasing detector mass or live time. Instead, sensitivity enhancement critically depends on sustained reduction of background index, improved energy resolution, and enhanced event reconstruction techniques.

Using the standard light-Majorana-neutrino exchange framework, the extracted effective Majorana mass ranges indicate that present-generation xenon experiments probe predominantly the IH region and the quasi-degenerate neutrino mass regime. In contrast, the NH region remains experimentally inaccessible because destructive Majorana phase interference suppresses the effective mass to the few-meV scale.

A major aspect of the present analysis is the explicit consideration of FSI related systematic uncertainties in the interpretation of the effective Majorana mass. These effects broaden the experimentally inferred \(\langle m_{\beta\beta} \rangle\) intervals and influence the hierarchy sensitivity differently in the NH and IH regimes. In the NH region, where the effective mass is intrinsically small,
\[\langle m_{\beta\beta}\rangle_{\rm NH}\sim 1.5\text{--}3.7~\mathrm{meV},\]
even modest reconstruction uncertainties produce large fractional variations in the extracted sensitivity. In contrast, the IH region,
\[\langle m_{\beta\beta}\rangle_{\rm IH}\sim 18.7\text{--}49.2~\mathrm{meV},\]experiences larger absolute uncertainty in meV scale, although the corresponding fractional uncertainty remains comparatively moderate due to the larger signal magnitude.

The exposure estimates further demonstrate the strong scaling relation
\[\Sigma_{\rm req}\propto m_{\beta\beta}^{-4},\]
which causes the required detector exposure to increase rapidly as the target effective mass decreases. While next-generation ton-scale detectors such as nEXO may fully probe the IH band, accessing the NH region would require exposures several orders of magnitude larger together with unprecedented control of background fluctuations, detector systematics, energy calibration, and FSI-related reconstruction effects.

The present comparison should nevertheless be interpreted carefully. Some sensitivities are derived from projected detector configurations, ROI definitions differ across experiments, and the extracted mass ranges remain dependent on the adopted NME calculations. Consequently, the primary value of this work lies not in establishing a strict ranking among experiments, but in providing a transparent statistical framework for consistent cross-technology comparison and hierarchy interpretation.

\subsection{Novelty of the Present Work}

The present work contains several features that distinguish it from conventional \(0\nu\beta\beta\) sensitivity studies:

\begin{itemize}

\item[i] \textbf{Unified Exposure-Based Statistical Framework:}  
A common isotope-exposure normalization framework was developed to compare xenon-based experiments consistently across different detector technologies, exposures, and ROI definitions.

\item[ii] \textbf{Direct Background Normalization:}  
Instead of relying only on published half-life sensitivities, the analysis explicitly computes expected background counts through
\[N_b = \mathrm{BI}\,\Delta E\,\Sigma_{\mathrm{iso}},\]
providing a transparent connection between detector design parameters and statistical sensitivity.
\item[iii]\textbf{Cross-Technology Comparison:}  
The work compares liquid scintillator, gaseous TPC, and dual-phase xenon detector technologies within a common statistical formalism, enabling a more uniform interpretation of detector performance.
\item [iv]\textbf{Hierarchy-Oriented Mass Interpretation:}  
The extracted effective Majorana mass sensitivities are directly interpreted in terms of the normal-hierarchy and inverted-hierarchy parameter spaces using oscillation data.
\item[v] \textbf{Inclusion of FSI/Reconstruction Effects:}  
Unlike many simplified sensitivity analyses, the present work explicitly discusses the impact of final-state interaction (FSI) and reconstruction-related systematic uncertainties on the extracted effective Majorana mass ranges and hierarchy reach.
\item[vi] \textbf{Exposure Scaling Toward NH Sensitivity:}  
The study quantitatively estimates the exposure required to probe representative NH and IH mass scales, demonstrating the severe experimental challenge associated with reaching the few-meV NH regime.
\item[vii] \textbf{Detector Optimization Perspective:}  
The framework not only compares existing experiments but also identifies the critical detector parameters governing future sensitivity, including ROI width, energy resolution, topology discrimination, and systematic uncertainty control.
\end{itemize}

\section{Conclusion}

In this work, we developed a unified statistical framework for comparing xenon-based neutrinoless double beta decay searches in \(^{136}\)Xe using isotope exposure, ROI width, background index, and effective Majorana mass sensitivity within a common formalism. The framework provides a transparent method for evaluating detector performance in the low-count Poisson regime relevant for rare-event searches.
Using a matched isotope exposure of
\[\Sigma_{\mathrm{iso}} = 100~\mathrm{kg\cdot yr},\]
the expected background ordering obtained in this study is
\[N_b^{\mathrm{nEXO}}<  N_b^{\mathrm{NEXT}}<N_b^{\mathrm{KLZ}}<N_b^{\mathrm{LZ}}\]
demonstrating that experimental sensitivity is governed primarily by background suppression capability and energy resolution rather than detector mass alone.

The analysis further clarifies the transition between background-free and background-dominated sensitivity regimes, emphasizing that future improvements in \(0\nu\beta\beta\) searches require simultaneous progress in detector exposure, background reduction, and event reconstruction performance.

Interpreted within the standard light-Majorana-neutrino exchange framework, the present-generation xenon experiments primarily probe the inverted-hierarchy region and the quasi-degenerate neutrino mass regime, while the normal-hierarchy region remains experimentally inaccessible because of the strong suppression of the effective Majorana mass due to Majorana phase cancellations.

A central result of this work is the demonstration that FSI and reconstruction-related uncertainties play an increasingly important role as experiments approach the few-meV sensitivity scale. In the NH region, even small reconstruction effects generate large fractional uncertainties because the effective mass itself is strongly suppressed. Consequently, future NH-sensitive experiments will require not only multi-ton exposures but also exceptional control of detector systematics, energy calibration, ROI optimization, and background discrimination.

The exposure estimates obtained here indicate that next-generation experiments such as nEXO may achieve comprehensive sensitivity across the IH parameter space, whereas probing the complete NH region will remain one of the major long-term challenges in experimental neutrino physics.

Overall, the framework developed in this work provides a consistent statistical basis for cross-experimental comparison, hierarchy interpretation, detector optimization, and future sensitivity projections for xenon-based \(0\nu\beta\beta\) searches.  
\bibliographystyle{unsrt}  
\bibliography{references}   
\end{document}